\documentclass[a4paper,12pt]{article}
\usepackage[a4paper,text={16.8cm,22.8cm}]{geometry} 
\usepackage{amsmath,amssymb,bm,graphicx}
\usepackage{mathtools}
\usepackage{mathrsfs} 
\usepackage[dvipsnames]{xcolor}
\usepackage{extarrows}
\usepackage[utf8]{inputenc}
\usepackage{amsthm}
\usepackage{slashed}
\usepackage{bm}
\usepackage{rotating}
\usepackage{hyperref}
\usepackage{array}
\usepackage{multirow}
\usepackage{bbold}
\usepackage{cite}
\usepackage{changepage}
\usepackage{pifont}
\usepackage[]{relsize}
\usepackage{tabularx}
\usepackage{stackengine}
\usepackage{gensymb}
\usepackage[]{cancel} 
\usepackage{caption}
\usepackage{subcaption}
\usepackage{soul} 
\usepackage{textcomp}

\allowdisplaybreaks 
\usepackage[export]{adjustbox}
\usepackage{float}
\theoremstyle{definition}

\let\originalleft\left
\let\originalright\right
\renewcommand{\left}{\mathopen{}\mathclose\bgroup\originalleft}
\renewcommand{\right}{\aftergroup\egroup\originalright}

\newcommand\be{\begin{equation}}
\newcommand\ee{\end{equation}}

\begin{document}

\begin{titlepage}

\begin{flushright} 
LA-UR-25-31854
\end{flushright}

\vspace{1.2cm}
\begin{center}
\large\bf
\boldmath
Residual Symmetries and Scalar Multiplet Vacuum Alignment in Non-Abelian Flavour Models
\unboldmath
\end{center}
\vspace{0.2cm}
\begin{center}
Ivo de Medeiros Varzielas,$^{a}$ Ming-Shau Liu,$^{b}$ Amartya Sengupta,$^{c,d}$ and Jim Talbert$^{e,f}$\\
\vspace{1.0cm}
{\small\sl 
${}^a$\,CFTP, Departamento de F\'{i}sica, Instituto Superior T\'{e}cnico, Universidade de Lisboa, Avenida Rovisco Pais 1, 1049 Lisboa, Portugal,\\[0.1cm]
${}^b$\,Homerton College, University of Cambridge, Hills Road, Cambridge, CB2 1PH, United Kingdom\\[0.1cm]
${}^c$\,Department of Physics, SUNY at Buffalo, Buffalo, NY 14260, USA\\[0.1cm]
${}^d$Department of Physics, University of Cincinnati, Cincinnati, Ohio 45221, USA\\[0.1cm]
${}^e$\,Institute for Defense Analyses, 730 E. Glebe Rd., Alexandria, VA  22305, USA\\[0.1cm]
${}^f$\,Theoretical Division, Group T-2, MS B283, Los Alamos National Laboratory, P.O. Box 1663, \\Los Alamos, NM  87545, USA}\\[0.5cm]
{\bf{E-mail}}: ivo.de@udo.edu, msl63@cam.ac.uk, amartyas@buffalo.edu, rjt89@cantab.ac.uk\\[1.0cm]
\end{center}

\vspace{0.5cm}
\begin{abstract}
\vspace{0.2cm}
\noindent 
We demonstrate that, upon minimizing a renormalizable, single-scalar potential invariant under a non-Abelian symmetry, special orientations in the associated vacuum alignment of the scalar multiplet correspond to the preservation of a discrete residual flavour symmetry in the broken phase of the theory.  Conversely, we show that these special scalar alignments are perturbed when additional Lagrangian operators (e.g. renormalizable, multi-flavon operators and/or effective, higher-dimensional operators) are present that break said residual symmetry, leading to a vacuum reorientation and phenomenological consequences.  We therefore construct a one-to-one correspondence principle between broken residual symmetries and vacuum alignment corrections, providing a mechanism to identify (and correct) a subtle but persistent form of phenomenologically relevant fine-tuning embedded in --- but often ignored by --- most successful non-Abelian flavour models.  We first establish this correspondence in a set of toy models based on the S4 permutation symmetry, and then apply the lessons learned to the more realistic A4 Altarelli-Feruglio and $\Delta(27)$ Universal Texture Zero models. 

\end{abstract}
\vfil

\end{titlepage}


\tableofcontents
\noindent \makebox[\linewidth]{\rule{16.8cm}{.4pt}}


\section{Introduction and Motivation}
\label{sec:INTRO}

The pursuit of a convincing theory explaining the patterns of fermionic mass, mixing, and CP violation observed in Standard Model (SM) processes is both longstanding and ongoing.  Popular solutions to this so-called \emph{flavour puzzle} employ a Beyond-the-SM (BSM) symmetry $\mathcal{G}_F$, valid at energy scales $\Lambda_F \gg \Lambda_{\text{SM}}$, under which different generations of SM particles and exotic scalar \emph{flavon}
degrees of freedom are charged.  Flavon(s) form a scalar potential with the Higgs boson $H$, and its minimization in flavour space yields a vacuum expectation value (VEV) that breaks $\mathcal{G}_F$, as required to realize the dramatic mass hierarchies seen in Nature. 

The specific structure of a `flavoured' scalar potential depends on the particle and symmetry content of a given model. 
For example, Froggatt-Nielsen theories \cite{Froggatt:1978nt} employ a single flavon and an Abelian U(1) flavour symmetry to resolve each matrix element of the SM Yukawa couplings in terms of a unique effective field theory (EFT) operator.  If $\mathcal{G}_F$ is instead non-Abelian, different flavours can sit in a single, multi-dimensional irreducible representation of the symmetry group --- a `multiplet.'  Unlike U(1) Froggatt-Nielsen theories, non-Abelian models tend to introduce multiple flavons with distinct contractions amongst SM fermions.

Both continuous and discrete non-Abelian symmetry groups are popular in the model-building literature.  The unified SU(3) theories of King, Ross, and d.M. Varzielas \cite{King:2001uz,King:2003rf, deMedeirosVarzielas:2005ax} and the more recent SO(3) model of Reig-Valle-Wilczek \cite{Reig:2018ocz} are interesting examples of the former, while the Altarelli-Feruglio \cite{Altarelli:2005yp,Altarelli:2005yx} and Babu-Ma-Valle \cite{Babu:2002dz} A4 models are famous examples of the latter.  Ubiquitous to all of these models is the presence of flavon multiplets whose VEVs are aligned along special directions in flavour space.  
For instance, the more recent Universal Texture Zero (UTZ) model \cite{deMedeirosVarzielas:2017sdv,Bernigaud:2022sgk} employs five flavons, $\theta$ and $\theta_{\lbrace 3,123,23,X\rbrace}$, charged as triplets under $\Delta(27)$ \cite{deMedeirosVarzielas:2006fc,Ma:2006ip,Luhn:2007uq,deMedeirosVarzielas:2015amz,Ishimori:2010au}, and whose VEVs are taken to be\footnote{We have set the complex phases permitted in these VEVs (cf. Appendix A in \cite{deMedeirosVarzielas:2017sdv}) to $\lbrace\alpha,\beta \rbrace = \pi$.}
\begin{equation}
\left\langle {{\theta _{(3)}}} \right\rangle  = {{\rm{v}}_{\theta (3)}}\left( {\begin{array}{*{20}{c}}
0\\
0\\
1
\end{array}} \right),\quad \left\langle {{\theta _{123}}} \right\rangle  = -\frac{{{\rm{v}}_{123}}}{{\sqrt 3 }}\left( {\begin{array}{*{20}{c}}
1\\
1 \\
{1}
\end{array}} \right), \quad
 \left\langle {{\theta _{23}}} \right\rangle  = \frac{{{\rm{v}}_{23}}}{{\sqrt 2 }}\left( {\begin{array}{*{20}{c}}
0\\
-1\\
1
\end{array}} \right),\quad \left\langle {{\theta ^{\dag}_X}} \right\rangle  = \frac{{{\rm{v}}_X}}{{\sqrt 6 }}\left( {\begin{array}{*{20}{c}}
-2\\
1 \\
1
\end{array}} \right)\,,
\label{eq:UTZvevs}
\end{equation}
where v$_{\lbrace 3,123,23,X \rbrace}$ are scales, and the first equality indicates that the alignment is the same for both $\theta$ and $\theta_3$.  \eqref{eq:UTZvevs} is not unique to the UTZ; one or more of these VEV orientations appear in all of the models mentioned above, up to minus signs.

The directions in \eqref{eq:UTZvevs} are found by minimizing a scalar potential in flavon field space, typically written down at the renormalizable level.  For example, the $\langle \theta_{3,123} \rangle$ directions can be achieved with a $\Delta(27)$-invariant potential given by
\begin{equation}
\label{eq:UTZsingleflavonV}
V_{4,\theta}= -m_\theta^2\left[\theta^\dagger\theta\right]_{\bm{1_{0,0}}}+\sum_i\lambda_{\theta i}\,\left[\theta^\dagger\theta^\dagger\theta\theta\right]_{\bm{1_{0,0}}},\quad\quad i=1\dots4,
\end{equation}
where the range of the index $i\le 4$ was deduced from a Hilbert series calculation with {\tt{DECO}} \cite{Rahn:2020eco,Rahn:2023deco},\footnote{We had to augment {\tt{DECO}} 's range of supported symmetries to include $\Delta(27)$.  We thank Rudi Rahn with his support in this effort.} $m_\theta$ is the bare mass of $\theta$, $\lambda_{\theta_i}$ is an arbitrary dimensionless coupling, and the ${\bf{1_{0,0}}}$ notation indicates a contraction to a trivial $\Delta(27)$ singlet.  Note that in SU(3) theories there is only one quartic coupling, and the corresponding potential does \emph{not} realize $\langle \theta_{3,123} \rangle$ as minima.\footnote{The ability to align vacua as in \eqref{eq:UTZvevs} has long been argued as motivation for using non-Abelian discrete symmetries -- cf. the reviews in \cite{Altarelli:2010gt,Ishimori:2010au,King:2013eh}.}  

However, \eqref{eq:UTZsingleflavonV} omits mixed operators composed of both $\theta_{3,123}$, assuming they are either absent or subdominant to single-flavon operators, despite being allowed by the model's symmetries. In the absence of some other dynamical mechanism, this assumption amounts to an arbitrary fine-tuning of the scalar potential.  $V_{4,\theta}$ is also written down at the renormalizable $d=4$ level, despite the fact that VEVs as in \eqref{eq:UTZvevs} are often `plugged in' to Yukawa sector operators written down at non-renormalizable $d>4$ orders --- cf. e.g. \cite{deMedeirosVarzielas:2017sdv, Bernigaud:2022sgk}.  This could lead to inconsistencies in the EFT power counting.  It is also well known that minimizing EFT scalar potentials at $d>4$ sources relevant effects in other Lagrangian sectors, as seen for example in \cite{Alonso:2013hga} for the case of non-renormalizable Higgs operators in the Standard Model Effective Field Theory (SMEFT) \cite{Buchmuller:1985jz,Grzadkowski:2010es}, or in (e.g.) \cite{Loisa:2024xuk} where the $d=6$ flavon potential in Froggatt-Nielsen EFTs yields a redefined flavon VEV and mass term.  

The critical point is that, due to the presence of additional operators fully allowed by the field and symmetry content of the model --- but otherwise frequently ignored in flavour models --- vacuum alignments as in \eqref{eq:UTZvevs} may be subject to corrections that reorient the scale and direction of the flavour symmetry breaking.  For example, $\langle \theta_3 \rangle$ may be corrected such that 
\begin{equation}
\langle \theta_3 \rangle  \longrightarrow  \langle \theta_3 \rangle^{\prime} = \left(
\begin{array}{c}
\delta_{\theta_1} \\
\delta_{\theta_2} \\
v_{\theta_3} + \delta_{\theta_3} 
\end{array}
\right)\,, 
\label{eq:vevcorrect}
\end{equation}
and similarly for other alignments.
These $\delta_{\theta_i}$ corrections can lead to phenomenologically-relevant effects in IR Yukawa sector predictions (and beyond) that are of equal or greater order to corrections coming from higher-order Yukawa EFT operators.  A priori, these vacuum alignment corrections might even re-motivate the overall symmetry and particle content of the model itself, since these are often designed to allow certain Yukawa operators and not others based on phenomenological considerations.

We aim to systematically understand when special vacuum alignments like those in \eqref{eq:UTZvevs} are stable against reorienting corrections $\delta$ as in \eqref{eq:vevcorrect}, in the absence of ad hoc fine-tuning (e.g. setting coefficients of mixed-flavon operators to zero).  This paper will show that special vacuum alignments correspond to the presence of residual flavour symmetries (RFS) $\mathcal{G}_{\text{RFS}}$ in the broken phase of the \emph{single-flavon} sector of the theory, and that are mediated by subgroups of the original flavour symmetry $\mathcal{G}_F$.  These RFS are typically discrete and can be Abelian or non-Abelian, such that the action of their generator(s) $\mathbb{G}_{i}$ on the VEV respect
\begin{equation}
\label{eq:RFSrelation}
\mathbb{G}_{i} \cdot \langle \theta \rangle \overset{!}{=}\langle \theta \rangle \, \quad \forall i 
\end{equation}
for an arbitrary flavon multiplet $\theta$.  The invariance in \eqref{eq:RFSrelation} is analogous to the RFS frequently studied in Yukawa-like sectors \cite{Lam:2007qc,Hernandez:2012ra,Lam:2012ga,Holthausen:2012wt,King:2013vna,Holthausen:2013vba,Lavoura:2014kwa,Fonseca:2014koa,Joshipura:2014qaa,Talbert:2014bda,Yao:2015dwa,Lu:2016jit,deMedeirosVarzielas:2016fqq,deMedeirosVarzielas:2019lgb,Varzielas:2023qlb,Desai:2023jxh,Bernigaud:2019bfy,Bernigaud:2020wvn,deMedeirosVarzielas:2019dyu}.\footnote{In the nomenclature of \cite{King:2013eh}, the relationship in \eqref{eq:RFSrelation} would correspond to a `direct' RFS model.}  When models incorporate multiple flavons $\theta_{A,B,C,..}$, there are distinct RFS corresponding to each independent VEV, and the overall RFS is given by
\begin{equation}
\text{RFS} \sim \mathcal{G}^{\theta_A}_{N_A} \times \mathcal{G}^{\theta_B}_{N_B}  \times \mathcal{G}^{\theta_C}_{N_C}   \times \quad \text{...}\,,
\end{equation}
where $N_{A,B,C}$ denote the orders of the subgroups.

Our analysis will show that operators that respect the identified RFS (upon expanding about the simultaneous minima defined by the VEVs) preserve the VEV orientations derived from the single-flavon potentials, while those that break the RFS lead to non-trivial reorientations in flavour space.  In other words, we identify a one-to-one correspondence principle between broken scalar RFS and VEV realignment in the absence of Lagrangian coupling fine-tunings.  This correspondence represents a mechanism for identifying when --- and how --- predictions made at the level of IR Yukawa textures are robust against fine-tuning in the scalar sector.  We further demonstrate a perturbative method used in \cite{Altarelli:2005yx} to analytically calculate the VEV corrections $\delta$ induced by the RFS-violating operators, which can alleviate some fine-tuning and alter the flavour predictions of the theory.

Our work adds to surprisingly little research on the topic. In one of the most detailed analyses of this kind, Altarelli and Feruglio engaged in a  thorough calculation of power-suppressed EFT contributions to vacuum alignment in \cite{Altarelli:2005yx}, showing that they could reorient the leading-order A4 vacuum non-trivially.  We will address this in detail in Section \ref{sec:A4AF}.  Additionally, the authors of \cite{deMedeirosVarzielas:2012ylr} examined certain $\Delta(3\cdot n^2)$-invariant non-renormalizable scalar operators formed from multiple Higgs doublets, showing that the minima of potentials including them could still preserve `geometrical' (model-parameter-free) CP violation predictions.  Similarly, \cite{Alonso:2011yg} analyzed the non-renormalizable scalar potential of spurions (flavons) assuming Minimal Flavor Violation (MFV) \cite{DAmbrosio:2002vsn}, asking whether its minimization could achieve realistic Yukawa textures when said flavons transformed in the bi-fundamental ($d=5$) and/or fundamental ($d=6$) representations of the MFV quark symmetry group, SU(3)$_{Q_L} \times$ SU(3)$_{U_R} \times$ SU(3)$_{D_R}$.  In \cite{Hagedorn:2023mrg}, the effect of supersymmetric (SUSY) soft-breaking terms on alignments obtained from SUSY theories is analysed, and general conditions for the preservation of the alignments are presented, analogously to those found for multi-Higgs models in \cite{deMedeirosVarzielas:2021zqs}.

An important solution to the problem of vacuum alignment is proposed in \cite{Babu:2010ex}, where an enlarged flavour group is employed. In that model the explicit flavour symmetry is (S3)$^4 \rtimes$ A4, supplemented by an additional $\mathbb{Z}_2$ \cite{Babu:2010ex}. It is possible to forbid unwanted operators leading to flavon realignments via this enlarged symmetry.
This idea was subsequently developed into a more group-theoretic framework in \cite{Holthausen:2011pd}.  There, for a given flavour group $\mathcal{G}_F$
with a non-trivial extension $\mathcal{N}$, the scalar potential respects an enlarged symmetry $\mathcal{G}^\prime_F$ defined by
\begin{equation}
\mathcal{G}^\prime_F = \mathcal{N} \rtimes \mathcal{G}_F\,, \qquad \mathcal{G}^\prime_F/\mathcal{N} \simeq \mathcal{G}_F\,.
\end{equation}
The scalar potential is then left with a larger symmetry than the effective flavour symmetry of terms involving the fermions such that, at the renormalizable level, two flavons ($\theta$, $\phi$) only couple in the potential through the trivial singlet contraction:
\begin{equation}
\label{eq:V4mixBigGroup}
V_{4}(\theta,\phi) = V_{4}(\theta) + V_{4}(\phi) + \gamma\,(\theta\theta)_{\mathbf{1}}(\phi\phi)_{\mathbf{1}}\,,
\end{equation}
where other mixed-flavon contractions are forbidden.
\cite{Holthausen:2011pd} presents a minimal Q8 $\rtimes$ A4 example.

The point of the present work is different, although closely related in spirit. 
We do not attempt to construct a new enlarged-group alignment mechanism of the type above. 
Instead, we present a general symmetry-based criterion that diagnoses when a proposed flavon vacuum alignment is protected in the broken phase of the theory or, alternatively, when it is expected to be destabilized via RFS-violating operators. 
Hence, from our point of view, the enlarged symmetry structures of \cite{Babu:2010ex,Holthausen:2011pd} work to protect vacuum alignments precisely because they forbid those mixed operators that would otherwise violate the RFS associated with the target VEVs.
Our approach therefore does not compete with these earlier mechanisms; rather, it provides a unifying principle for understanding why they succeed.
In this sense, our framework is more diagnostic than UV-complete: it tells one which operators are dangerous and therefore indirectly suggest what kind of enlarged symmetry, sequestering, or shaping structure would be required to eliminate them in order to preserve preferred vacuum alignments as in \cite{Babu:2010ex,Holthausen:2011pd}. We also calculate the associated misalignments in the limit of soft RFS breaking, which could have phenomenological and model-building implications.
The realistic examples studied below will illustrate this point explicitly.

The paper develops as follows:  in Section \ref{sec:S4} we formalize our correspondence principle in simple one- and two-flavon models based on the S4 permutation symmetry, including both mixed-multiplet and EFT analyses.  Then in Section \ref{sec:generalize} we extend our findings to two realistic models in the literature, the Altarelli-Feruglio and Universal Texture Zero models mentioned above.  We provide a concluding summary and outlook in Section \ref{sec:CONCLUDE}.

\section{S4 Toy Models}
\label{sec:S4}
We first study toy models based on one or two flavons, with scalar potentials invariant under S4, the smallest non-Abelian group in the $\Delta(6 n^2)$ group series ($n=2 \leftrightarrow S4$) with a triplet irreducible representation.\footnote{$\Delta(6) \sim \text{S3}$, the permutation group of the triangle, is the smallest non-Abelian member of the group series, but it only furnishes singlet and doublet irreducible representations.} S4 is the symmetric permutation group of 24 elements and corresponds to the symmetry group of a cube.  It has been studied in many contexts, and is a popular candidate for flavoured model building. Relevant details of its mathematical structure are found in 
(e.g.) \cite{Ishimori:2010au}. Of course, the points we make below could equally well be made when studying Lagrangians invariant under different non-Abelian groups, and in Section \ref{sec:generalize} we do just that.  In this Section, we aim simply to establish our correspondence principle between vacuum (re)alignment and RFS violation.  

\subsection{$d=4$ Single-Flavon Potential and RFS Identification}
\label{sec:LOV}
We first write down the S4-symmetric potential when only a single triplet flavon $\theta \sim \bf{3}$ is present in the theory:
\begin{equation}
\label{eq:V4}
V^4_\theta = -m_\theta^2 \left[\theta^\dagger \theta\right]_{\bf{1}}+ \lambda^\theta_i \sum_i \left[ \theta^\dagger \theta^\dagger \theta \theta \right]^i_{\bf{1}} \,,
\end{equation}
where the bracket notation indicates contraction to an S4 singlet ${\bf{1}}$, and the summation is over independent S4 invariants.  We have also assumed the presence of an additional Abelian shaping symmetry under which $\theta$ is charged non-trivially.  These symmetries are commonly employed in flavour models to limit interactions amongst distinct scalar flavons in Yukawa operators, allowing one to `shape' their associated Yukawa couplings in phenomenologically favorable ways.  At the level of the single-flavon S4 toy model potentials, they effectively limit us to operators with even mass dimension $d$. 

Utilization of group product rules or the {\tt (D)ECO} \cite{Rahn:2023deco} Hilbert-series generator shows that $V_\theta^4$ contains four $S_4$-invariant operators in total: one quadratic mass term and three independent quartic invariants. Accordingly, the quartic sum runs over i=1,2,3. These invariants correspond to the 
\begin{equation}
\lambda^\theta_1\,\left[\left[\bf{3}\bf{3}\right]_{\bf{1}}\left[\bf{3}\bf{3}\right]_{\bf{1}}\right]_{\bf{1}}\,,\,\,\,\,\, \lambda^\theta_2\,\left[\left[\bf{3}\bf{3}\right]_{\bf{2}}\left[\bf{3}\bf{3}\right]_{\bf{2}}\right]_{\bf{1}} \,,\,\,\,\,\, \text{and} \,\,\,\,\,\, \lambda^\theta_3\,\left[\left[\bf{3}\bf{3}\right]_{\bf{3}}\left[\bf{3}\bf{3}\right]_{\bf{3}}\right]_{\bf{1}}
\end{equation}
contractions.  The contraction in the flavon mass term is of course just $\left[\bf{3}\bf{3}\right]_{\bf{1}}$.  Assuming real triplet components ($\theta_i = \theta_i^\dagger$) 
, we show the decomposed S4-invariant potential explicitly:
\begin{align}
\label{eq:V4decompose}
V^4_\theta = &-m_\theta^2 \left(\theta_1^2  + \theta_2^2 + \theta_3^2 \right) + \left(\lambda^\theta_1 + \frac{2 \lambda^\theta_2}{3} \right) \left(\theta_1^4 + \theta_2^4 + \theta_3^4 \right)
+ \left(2 \lambda^\theta_1 - \frac{2 \lambda^\theta_2}{3} + 4 \lambda_3^\theta\right) \left(\theta_1^2 \theta_2^2 + \theta_1^2 \theta_3^2 + \theta_2^2 \theta_3^2 \right)
\end{align}
We reorganized the expression so that an analogue can be drawn to the invariants identified in \cite{King:2009ap}, accounting for our real-field assumptions.

As $V^4_\theta$ is S4-symmetric, it of course follows that it is invariant under the action of all 24 S4 group elements (including its generators).  We will now consider the potential when S4 is broken by particular VEVs $\langle \theta \rangle$ of the flavon.

\subsubsection{%
\texorpdfstring{$\langle\theta\rangle\sim(0,0,1)$}{<theta>~(0,0,1)} and its
\texorpdfstring{$\mathbb{Z}_2\times\mathbb{Z}_2$}{Z2 x Z2} RFS}
\label{sec:00v}

It is straightforward to minimize $V^4_\theta$ with respect to the field components $\theta_i$.  Doing so, identifying the Lagrangian constraints at the desired critical point, and studying the eigenvalues of the Hessian operator at said point, one sees that the $\langle \theta_3 \rangle \equiv  v_\theta \left(0,0,1\right)$ VEV is realized if
\begin{equation}
\label{eq:00vVEVcondition}
v_\theta = \frac{\sqrt{3}\,\,  m_\theta}{\sqrt{6 \lambda^\theta_1 + 4 \lambda^\theta_2}}\,\,\,\,\,\text{and}\,\,\,\,\, \frac{3 \lambda^\theta_3 }{3 \lambda^\theta_1 + 2 \lambda^\theta_2} > 0\,.
\end{equation}
At this minimum the theory enters a phase where S4 is spontaneously broken, and our goal is to identify a RFS 
that is preserved in this phase.  To do so, we generate all 24 elements of S4 and then scan through them to determine those matrices whose action leaves $\langle \theta \rangle$ invariant.  It is straightforward to deduce that there are four group elements (including the identity) that do so, and that they can be generated by a further subset of two elements $T_{3,3b}$:
\begin{align}
\nonumber
T_3 \equiv G_2 \cdot G_1 \cdot G_2 \cdot G_1 &= \left(
\begin{array}{ccc}
-1 & 0 & 0 \\
0 & -1 & 0 \\
0 & 0 & 1
\end{array}
\right)\,\,\,\,\,\,\,\Leftrightarrow\,\,\,\,\,\,T_3 \langle \theta_3 \rangle = \langle \theta_3 \rangle\,, \\
\label{eq:T3}
T_{3b} \equiv G_1 \cdot G_2 \cdot G_2 &= \left(
\begin{array}{ccc}
0 & -1 & 0 \\
-1 & 0 & 0 \\
0 & 0 & 1
\end{array}
\right)\,\,\,\,\,\,\,\Leftrightarrow\,\,\,\,\,\,T_{3b} \langle \theta_3 \rangle = \langle \theta_3 \rangle \,.
\end{align}
Here $G_{1,2}$ are the generators of S4 in the triplet basis given by
\begin{equation}
\label{eq:S4tripgens}
G_1 = \left(
\begin{array}{ccc}
-1 & 0 & 0 \\
0 & 0 & -1 \\
0 & 1 & 0
\end{array}
\right),\,\,\,\,\,\,\,\,\,\,
G_2 =
\left(
\begin{array}{ccc}
0 & 0 & 1 \\
1 & 0 & 0 \\
0 & 1 & 0
\end{array}
\right)\,.
\end{equation}
$G_1$ generates a $\mathbb{Z}_4$ subgroup of S4 ($(G_1)^4 = \mathbb{1}$), while $G_2$ generates a $\mathbb{Z}_3$ subgroup ($(G_2)^3 = \mathbb{1}$). 

$T_3$ and $T_{3b}$ are elements of S4 that individually form Abelian $\mathbb{Z}_2$ subgroups, which upon closure with one another yield an Abelian product group.  Hence we identify our RFS as:
\begin{equation}
\text{RFS} \vert_{\langle \theta_3 \rangle} \sim \mathbb{Z}_2 \times \mathbb{Z}_2 \,.
\end{equation}
To test the invariance of the spontaneously broken potential, we must now expand it about the relevant vacuum,
\begin{equation}
\label{eq:00vexpand}
\langle \theta_3 \rangle \longrightarrow \left(\xi_{\theta _1},\xi_{\theta_2}, v_\theta + \xi_{\theta_3} \right)
\end{equation}
where $\xi_{\theta_i}$ are real field excitations.  Doing so results in a  broken-phase potential given by
\begin{align}
\nonumber
\slashed{V}^4_\theta \longrightarrow  &\left(2\lambda^\theta_1 - \frac{2\lambda^\theta_2}{3} +4 \lambda^\theta_3 \right)\left(\left(\xi_{\theta_1}^2 + \xi_{\theta_2}^2\right)\left(v_\theta + \xi_{\theta_3} \right)^2 + \xi_{\theta_1}^2 \xi_{\theta_2}^2\right) \\
\label{eq:VT3break}
&+ \left(\lambda_1^\theta + \frac{2 \lambda^\theta_2}{3}\right)\left(\xi_{\theta_1}^4 + \xi_{\theta_2}^4 + \left(v_\theta + \xi_{\theta_3}  \right)^4\right) - m_\theta^2 \left(\xi_{\theta_1}^2 + \xi_{\theta_2}^2 + \left(v_\theta + \xi_{\theta_3}  \right)^2 \right)  \,,
\end{align}
which is clearly invariant under the action of $T_{3,3b}$ and the remaining non-trivial $\mathbb{Z}_2 \times \mathbb{Z}_2$ group element, but (critically) \emph{not} that of other S4 elements, e.g. $G_{1,2}$. 
\subsubsection{%
\texorpdfstring{$\langle \theta \rangle \sim (0,-1,1)$}{<theta> ~(0,-1,1)} and its
\texorpdfstring{$\mathbb{Z}_2$}{Z2} RFS%
}
\label{sec:0mvv}
Following the procedure from \ref{sec:00v}, we find that 
\begin{equation}
v_\theta = \frac{\sqrt{3/2}\,m_\theta}{\sqrt{6 \lambda^\theta_1 + \lambda^\theta_2 + 6 \lambda^\theta_3}}\,\,\,\,\,\text{and}\,\,\,\,\, \frac{\lambda^\theta_{2,3}}{6 \lambda^\theta_1 + \lambda^\theta_2 + 6 \lambda^\theta_3} > 0\,
\end{equation}
realize the $\langle \theta_{23} \rangle \equiv  v_\theta \left(0,-1,1\right)$ VEV.  Searching for symmetry generator(s) that leave this VEV invariant leads one to identify 
\begin{equation}
\label{eq:T23}
T_{23} \equiv G_2^2 \cdot G_1^3 \cdot G_2^2 = \left(
\begin{array}{ccc}
1 & 0 & 0 \\
0 & 0 & -1 \\
0 & -1 & 0
\end{array}
\right)\,\,\,\,\,\,\Leftrightarrow \,\,\,\,\, T_{23} \langle \theta_{23} \rangle = \langle \theta_{23} \rangle 
\end{equation}
as the candidate RFS --- no other S4 elements leave $\langle \theta_{23} \rangle$ invariant.   $T_{23}$ generates an Abelian $\mathbb{Z}_2$ subgroup of S4, $T_{23}\cdot T_{23} \equiv \mathbb{1}$: 
\begin{equation}
\text{RFS} \vert_{\langle \theta_{23} \rangle} \sim \mathbb{Z}_2 \,.
\end{equation}
Expanding about this particular minimum via
\begin{equation}
\langle \theta_{23} \rangle \longrightarrow \left(\xi_{\theta_1}, -v_\theta + \xi_{\theta_2}, v_\theta + \xi_{\theta_3} \right)
\end{equation}
leads to 
\begin{align}
\nonumber
\slashed{V}^4_\theta \,\,\,\longrightarrow \,\,\,&\left(\chi_{+}^4 + \xi_{\theta_1}^4\right) \left(\lambda^\theta_1 + \frac{2\lambda^\theta_2}{3}\right) + 2\chi_{+}^2 \xi_{\theta_1}^2 \left(\lambda^\theta_1 - \frac{ \lambda^\theta_2}{3} + 2 \lambda^\theta_3- \frac{m_\theta^2}{2\,\xi_{\theta_1}^2}\right) \\
\label{eq:VT23break}
&+ \zeta\left[-\xi_{\theta_2},\xi_{\theta_3}\right]\left(2 \lambda^\theta_1 - \frac{2\lambda^\theta_2}{3} +4 \lambda^\theta_3\right) - m_\theta^2\, \xi_{\theta_1}^2\,,
\end{align}
where we defined 
\begin{align}
\chi^x_{\pm} &\equiv \left(\left(\xi_{\theta_2} - v_\theta \right)^x \pm \left(\xi_{\theta_3} + v_\theta \right)^x\right)\,,\\
\zeta[\alpha,\beta] &\equiv \left(v_\theta + \alpha\right)^2\left(v_\theta + \beta\right)^2\,,
\end{align}
for compactness.  The action of $T_{23}$ on $\xi_{\theta_{1,2,3}}$ in \eqref{eq:VT23break} leaves $\slashed{V}^4_\theta$ invariant --- $T_{23} \, [\slashed{V}^4_\theta] = \slashed{V}^4_\theta$ --- whereas other symmetry generators do not.  
\subsubsection{%
\texorpdfstring{$\langle \theta \rangle \sim (1,1,1)$}{<theta> ~(1,1,1)} and its
\texorpdfstring{$\text{S3}$}{S3} RFS%
}
\label{sec:vvv}
Finally, the $\langle \theta_{123} \rangle \equiv  v_\theta \left(1,1,1\right)$ VEV is realized if
\begin{equation}
\label{eq:vvvVEVcondition}
v_\theta = \frac{m_\theta}{\sqrt{6 \lambda^\theta_1 + 8 \lambda^\theta_3}}\,\,\,\,\,\text{and}\,\,\,\,\, \frac{\lambda^\theta_2 + \lambda^\theta_3}{3 \lambda^\theta_1 + 4 \lambda^\theta_3} > 0\,.
\end{equation}
 Unlike the analyses in Sections \ref{sec:00v} and \ref{sec:0mvv}, this minimum features a non-zero VEV in each flavour direction.  Proceeding as before, we immediately note that the action of the $G_2$ generator in \eqref{eq:S4tripgens} leaves this VEV unchanged, $G_2 \langle \theta_{123} \rangle = \langle \theta_{123} \rangle$.  Performing a complete scan of S4 group elements, we actually note that six S4 elements (including the identity) do the job, of which three are order two elements, and two are order three elements.  All six elements can be generated by one $\mathbb{Z}_2$ and one $\mathbb{Z}_3$ element; we identify the latter as $G_2$, and call the former $T_{123}$.  It is given by
 \begin{equation}
\label{eq:T123}
T_{123} \equiv G_{1} \cdot G_2 \cdot G_1 \cdot G_1= \left(
\begin{array}{ccc}
0 & 0 & 1 \\
0 & 1 & 0 \\
1 & 0 & 0
\end{array}
\right)\,\,\,\,\,\,\Leftrightarrow \,\,\,\,\, T_{123} \langle \theta_{123} \rangle = \langle \theta_{123} \rangle\,. 
\end{equation}
Hence the RFS is that group closed by the combined action of $T_{123}$ and $G_2$, which amounts to the (non-Abelian) $S3$ permutation symmetry:
\begin{equation}
\text{RFS} \vert_{\langle \theta_{123} \rangle} \sim \text{S3} \,.
\end{equation}
Expanding about the $\langle \theta_{123} \rangle$ minimum via
\begin{equation}
\label{eq:vvvexpand}
\langle \theta_{123} \rangle \longrightarrow (v_\theta + \xi_{\theta_1}, v_\theta + \xi_{\theta_2}, v_\theta + \xi_{\theta_3})
\end{equation}
leads to a broken-phase potential of the following form:
\begin{align}
\label{eq:V4slashvvv}
\nonumber
\slashed{V}^4_\theta \rightarrow \,\,\,&\lambda_1^\theta\,\chi[1,1,1,2] -m_\theta^2\,\chi[1,1,1,1] + \frac{\lambda_2^\theta}{6} \, \left(3 \,\chi[0,1,-1,2] +\chi[-2,1,1,2]\right) \\
&+ 4\lambda_3^\theta \,\left(\zeta[\xi_{\theta_1},\xi_{\theta_2}] + \zeta[\xi_{\theta_1},\xi_{\theta_3}] + \zeta[\xi_{\theta_2},\xi_{\theta_3}] \right)\,, 
\end{align}
where we defined 
\begin{align}
\label{eq:chizetadefs}
\chi[a,b,c,d] &\equiv \left(a \left(v_\theta + \xi_{\theta_1} \right)^2 + b \left(v_\theta + \xi_{\theta_2} \right)^2 + c \left(v_\theta + \xi_{\theta_3} \right)^2 \right)^d \,.
\end{align}
As before, we check how \eqref{eq:V4slashvvv} transforms under the action of the S3 generators, finding complete invariance as expected. 

\subsection{Operators Beyond the $d=4$ Single-Flavon Potential}
\label{sec:beyondS4renorm}
We now consider three types of extensions to the renormalizable, single-flavon potential in \eqref{eq:V4}:  single-flavon non-renormalizable operators with mass dimension $d>4$, as would appear in EFTs (Section  \ref{sec:d6V}), as well as a pair of two-flavon theories (one triplet-singlet model in Section \ref{sec:singtripV}, and one triplet-triplet model in Section  \ref{sec:twotripV}) at the renormalizable level. 
We will see that introducing additional operators into the theory beyond those in \eqref{eq:V4} which \emph{break} the RFS invariance identified for the different alignments in Sections \ref{sec:00v}-\ref{sec:vvv} leads to non-zero perturbations away from the zeroes in $\langle \theta_i \rangle$, while introducing additional operators that respect said RFS do not --- they simply rescale existing non-zero elements.

\subsubsection{$d=6$ Effective Single-Flavon Potential}
\label{sec:d6V}  

Given that many flavour models are written down as EFTs in the Yukawa sector, it may be necessary to study the flavon potential at higher mass dimensions ($d>4$) as well in order to be consistent with EFT operator power counting.  To that end, we write down the single-flavon Lagrangian at $d=6$ which, thanks to the presence of the additional Abelian shaping symmetry, is the next-to-leading order operator set:
\begin{equation}
\label{eq:V6}
V^6_\theta = 
\frac{\rho^{\theta}_i}{\Lambda^2} \, \sum_i \left[\theta \theta \theta \theta^\dagger \theta^\dagger \theta^\dagger \right]^i_{\bf{1}}\,.
\end{equation}
Here $\Lambda$ is an unspecified ultraviolet (UV) energy scale where additional particle degrees of freedom can propagate, and which sets the size of the EFT expansion.  Again using {\tt{(D)ECO}} to calculate the Hilbert series, we find that there are six additional operators arising in the potential at $d=6$, which we identify as the 
\begin{align}
\nonumber
&\,\rho^\theta_1\,\left[\left[ \bf{3 3} \right]_{\bf{1}}\left[ \bf{3 3} \right]_{\bf{1}}\left[ \bf{3 3} \right]_{\bf{1}}\right]_{\bf{1}}, \,\,\,\,\,\,\,\,\,\,\,\,\,\,\,\,\,\,\, \rho^\theta_2\, \left[\left[ \bf{3 3} \right]_{\bf{1}}\left[\left[ \bf{3 3} \right]_{\bf{2}}\left[ \bf{3 3} \right]_{\bf{2}}\right]_{\bf{1}}\right]_{\bf{1}},\,\,\,\,\,\,\,\,\,\,\,\, \rho^\theta_3 \,\left[\left[ \bf{3 3} \right]_{\bf{1}}\left[\left[ \bf{3 3} \right]_{\bf{3}}\left[ \bf{3 3} \right]_{\bf{3}}\right]_{\bf{1}}\right]_{\bf{1}},\\
\label{eq:V6contractions}
& \rho^\theta_4\,\left[\left[ \bf{3 3} \right]_{\bf{2}}\left[\left[ \bf{3 3} \right]_{\bf{2}}\left[ \bf{3 3} \right]_{\bf{2}}\right]_{\bf{2}}\right]_{\bf{1}},\,\,\,\,\,\,\,\,\,\,\,\, \rho^\theta_5\,\left[\left[ \bf{3 3} \right]_{\bf{2}}\left[\left[ \bf{3 3} \right]_{\bf{3}}\left[ \bf{3 3} \right]_{\bf{3}}\right]_{\bf{2}}\right]_{\bf{1}},\,\,\,\,\,\,\,\,\,\,\,\, \rho^\theta_6\,\left[\left[ \bf{3 3} \right]_{\bf{3}}\left[\left[ \bf{3 3} \right]_{\bf{3}}\left[ \bf{3 3} \right]_{\bf{3}}\right]_{\bf{3}}\right]_{\bf{1}},
\end{align} 
S4-invariant contractions, using group product rules.  For brevity we will not write down the explicit form of the decomposed $d=6$ Lagrangian, as in \eqref{eq:V4decompose}, but it is straightforward to calculate with Clebsch-Gordan coefficients.  The complete single-flavon potential is then given by $V_\theta = V^4_\theta + V^6_\theta$.   

We have performed the same RFS analysis as above for the $\langle \theta_{3,23,123} \rangle$ VEVs derived from \eqref{eq:V6}.  In all instances we find that the RFS identified from the leading order $V_\theta^4$ is \emph{preserved} when including the additional Lagrangian operators derived from \eqref{eq:V6contractions}.  If the preservation of the RFS is in one-to-one correspondence with the direction of the VEV, then the leading-order vacuum alignments $\langle \theta_{3,23,123} \rangle$ should also be preserved, up to corrections to the non-zero vacuum elements.  To confirm this, we have calculated the corrective $\delta$ terms defined by\footnote{We omit the analogous calculation for $\langle \theta_{23} \rangle^\prime$  for brevity. Also, it is important to distinguish the VEV correction terms $\delta_{\theta_i}$ from the $\xi_{\theta_i}$ field excitations introduced to expand about the broken phase of the theory.}
\begin{equation}
\label{eq:VEVcorrect1}
\langle \theta_{3} \rangle^\prime \equiv \left(\delta_{\theta_1},\delta_{\theta_2},v_{\theta_3} + \delta_{\theta_3}\right)
,
\,\,\,\,\,\,\,\,\,\,\,\,\,\,\,\,\,\,\,\,\langle \theta_{123} \rangle^\prime \equiv v_{\theta_{123}}\, +\, \left(\delta_{\theta_{123_1}},\delta_{\theta_{123_2}},\delta_{\theta_{123_3}}\right),\end{equation}
using perturbative methods analogous to those employed in Appendix B of \cite{Altarelli:2005yx}, finding
\begin{align}
\nonumber
&\delta_{\theta_1} = \delta_{\theta_2} \simeq 0, \,\,\,\delta_{\theta_3} \simeq -\frac{9\,m_\theta^3}{16\,\Lambda^2}\frac{\left(3\sqrt{6}\, \rho^\theta_1 + 2\sqrt{6}\, \rho^\theta_2 + 4\, \rho^\theta_4\right)}{\left(3 \lambda^\theta_1 + 2\lambda^\theta_2\right)^{5/2}}\\
\label{eq:dim6VEVcorrections}
&\delta_{\theta_{123_1}} = \delta_{\theta_{123_2}} = \delta_{\theta_{123_3}} \simeq -\frac{3\, m_\theta^3}{8 \sqrt{2} \,\Lambda^2}\frac{\left(9\,\rho^\theta_1 + 12\, \rho^\theta_3 + 16\, \rho^\theta_6\right)}{\left(3 \,\lambda^\theta_1 + 4 \, \lambda^\theta_3\right)^{5/2}}\,,
\end{align}
 One observes that, as expected, there are no VEV corrections to null elements, and corrections to non-zero elements are universal, in that they shift the leading-order VEV by the same amount in each non-zero direction, preserving both the relative magnitude and directions of the leading-order vacua.  These statements are consistent with the idea that an RFS is present and being respected in the Lagrangian.
Finally, note that the $\delta$ terms in \eqref{eq:dim6VEVcorrections} are not fine-tuned as they are in \eqref{eq:twoflavondeltaperts} below, since the perturbative expansion is performed in the (physical) $\Lambda \rightarrow \infty$ limit.

\subsubsection{$d=4$ Triplet-Singlet Flavon Potential}
\label{sec:singtripV}
We now consider potentials when there are two flavons charged differently under S4. To that end, we take $\theta \sim {\bf{3}}$ and $\phi_s \sim {\bf{1^\prime}}$.  A Hilbert Series analysis with {\tt{DECO}} indicates that, in the presence of a shaping symmetry, only a single mixed $\phi-\theta$ operator is present.  Combined with the single-flavon terms, the renormalizable scalar potential is given by
\begin{equation}
\label{eq:tripsingV}
V_{\theta \phi_s}^4 = V_\theta^4 - m_\phi^2 \,\left[\phi_s^\dagger \phi_s\right]_{{\bf{1}}} + \lambda^\phi_1 \,\left[\phi_s^\dagger \phi_s\right]_{{\bf{1}}}^2 + \gamma_s \,\left[\phi_s^\dagger \phi_s\right]_{{\bf{1}}}\left[\theta^\dagger \theta\right]_{{\bf{1}}}\,,
\end{equation}
where the only non-trivial S4 contraction is the $\left[\bf{3}_\theta \bf{3}_\theta\right]_{\bf{1}}$ in the mixed term.  Note that this is equivalent to \eqref{eq:V4mixBigGroup} above, which we comment on below.

\subsubsection*{$\langle \theta \rangle \sim (0,0,1)$ and $\langle \phi_s \rangle \sim v_\phi$}

Since the VEV of the singlet is one-dimensional in flavour space, it is clear that there is no non-trivial action on it that will leave it invariant a l\'a \eqref{eq:RFSrelation}. However, following the RFS identification algorithm of the above sections is less straightforward in this scenario, as the character table for S4 reveals that there are order one (identity), two and three group elements that transform trivially (i.e., with phase 0) in the ${\bf{1}^\prime}$ representation of S4.  They are collected in the $C1$, $C3$, and $C8$ group conjugacy classes, respectively, in the notation of \cite{Ishimori:2010au}.   Hence one could legitimately identify $\mathbb{1}$, $\mathbb{Z}_2$, and $\mathbb{Z}_3$ as candidate RFS for $\langle \phi_s \rangle$. Since the practical action of all of these RFS is the same, we collectively denote them by $\mathbb{1}^\star$.

With this in mind, and taking the third family orientation ($\langle \theta \rangle = (0,0,v_\theta)$) as the triplet VEV,   the overall RFS for the broken-phase Lagrangian is then given by
\begin{equation}
\label{eq:tripsingrealRFS}
\text{RFS}\big\vert_{\langle \theta \rangle \times \langle \phi_s \rangle} \sim \left(\mathbb{Z}_2\times \mathbb{Z}_2\right)^\theta \times \mathbb{1}^{\star,\phi_s}\,,
\end{equation}
with the superscript indicating which flavon experiences (non-)trivial action under the RFS.  Here $\left(\mathbb{Z}_2\times \mathbb{Z}_2\right)^\theta$ is generated by \eqref{eq:T3}, as identified for $\langle \theta_3 \rangle$ in the single-flavon scenario.

We continue as before by breaking \eqref{eq:tripsingrealRFS} with $\langle \phi_s \rangle$ and $\langle \theta_3 \rangle$, and then expand the resulting expression using \eqref{eq:00vexpand} and $\langle \phi_s \rangle^\prime \rightarrow v_\phi + \xi_{\phi_s}$. We act on this expanded minimum with \eqref{eq:tripsingrealRFS}, and find that the RFS is  preserved. 
As in Section \ref{sec:twotripV}, this indicates that the joint limit $\langle \phi_s \rangle$, $\langle \theta_3 \rangle$ is a true minimum of the theory, and it is straightforward to deduce that the triplet VEV
expectation is given by\footnote{An exact expression for the singlet VEV $v_\phi$ can also be found.  Also, the shifting of the $\theta$ mass is the standard effect seen when coupling to a bilinear scalar term $\phi^\dagger \phi$; had we assumed a scale separation between $m_\theta$ and $m_\phi$, the right-hand-side of \eqref{eq:tripsingmassshift} would clearly indicate the need for (potentially) unnatural fine-tuning of the bare mass $m_\theta$.}
\begin{equation}
\label{eq:tripsingmassshift}
v_\theta = \frac{\sqrt{3}\, \overline{m}_\theta}{\sqrt{6\lambda^\theta_1 + 4 \lambda^\theta_2}}\,, 
\,\,\,\,\,\,\,\,\,\, 
\text{with}\,\,\,\,\,\,\overline{m}_\theta^2 = m_\theta^2 - v_\phi^2 \, \gamma_s  \,.
\end{equation}  
As a final check, we have again confirmed our results by calculating the hypothetical perturbative corrections as in \eqref{eq:VEVcorrect1} (defining $\langle \phi_s \rangle^\prime \equiv v_\phi + \delta_\phi$), finding that 
\begin{equation}
\delta_{\theta_1} = \delta_{\theta_2} = 0\,,
\end{equation}
while $\delta_{\theta_3}$ and $\delta_\phi$ are nonzero.  In other words, the leading-order single-flavon excitations are simply rescaled, but not realigned in flavour space. Note that this conclusion is consistent with the commentary around \eqref{eq:V4mixBigGroup}, where we discussed \cite{Holthausen:2011pd}'s approach to enlarging the overarching flavour symmetry to yield an `alignment-safe' multi-flavon potential as in \eqref{eq:tripsingV}.

The triplet-singlet toy model also highlights the importance of the invariance of the VEV itself, as opposed to the Lagrangian more broadly, when identifying the appropriate scalar RFS.  After all, it is clear that \eqref{eq:tripsingV} is invariant under a generic complex transformation of the singlet with phase $\alpha$, $\phi_s^\prime \longrightarrow \text{exp}(i \pi \alpha) \, \phi_s$, which reduces to an invariance under a sign flip when considering all possible S4 actions ($\phi_s \rightarrow \pm \,\phi_s$) on a non-trivial singlet ${\bf{1}^\prime}$, regardless of which minimum we choose to study for the triplet flavon $\theta$. Had we allowed for RFS action on $\langle \phi_s \rangle$ associated to S4 elements in the $C6$ or $C6^\prime$ group conjugacy classes \cite{Ishimori:2010au}, then $\xi_{\phi_s} \rightarrow - \xi_{\phi_s}$, the broken phase Lagrangian would \emph{not} have been invariant, and we would have wrongfully concluded that $\langle \theta \rangle$ and $\langle \phi_s \rangle$ could not be realized simultaneously without fine-tuning, which is inconsistent with a straightforward analysis of \eqref{eq:tripsingV}'s critical points and the calculation of the $\delta$ corrections mentioned above.  

\subsubsection{$d=4$ Two-Triplet Flavon Potential}
\label{sec:twotripV}
Another obvious extension to \eqref{eq:V4} is to consider the renormalizable potential when two triplets $\theta$ and $\phi$ are present. 
After all, it is easily intuited that the RFS identified for a particular VEV in the single-flavon cases of Section \ref{sec:singtripV} may not be preserved in the broken phase of theory assuming two special multiplet alignments, each of which may be associated to a different non-trivial RFS a priori.

Given two S4 triplets, $\theta,\phi \sim {\bf{3}}$, the Lagrangian of \eqref{eq:V4} is extended to 
\begin{equation}
\label{eq:twoflavonV}
V^4_{\theta \phi} = -m_\theta^2 \left[\theta^\dagger \theta\right]_{\bf{1}} - m_\phi^2 \left[\phi^\dagger \phi\right]_{\bf{1}} + \lambda^\theta_i \sum_i \left[ \theta^\dagger \theta^\dagger \theta \theta \right]^i_{\bf{1}} + \lambda^\phi_j \sum_j \left[ \phi^\dagger \phi^\dagger \phi \phi \right]^j_{\bf{1}} + \gamma_k \sum_k \left[ \theta^\dagger \phi^\dagger \theta \phi \right]^k_{\bf{1}} \,,
\end{equation}
where we assume that both $\theta$ and $\phi$ are distinctly charged under an Abelian shaping symmetry such that the cross-terms coupling to $\gamma_k$ have equal numbers of (un)conjugated $\theta$ and $\phi$ fields.  It is clear that, as in the case of \eqref{eq:V4}, $\lbrace i, j \rbrace \le 3$, whereas we use {\tt{(D)ECO}} to deduce that $k \le 4$. S4  group product rules and some algebra lead us to conclude that the
\begin{equation}
\gamma_1 \left[\left[\bf{3}_\phi\bf{3}_\phi\right]_{\bf{1}}\left[\bf{3}_\theta\bf{3}_\theta\right]_{\bf{1}}\right]_{\bf{1}},\,\,\,\,\,\, \gamma_2\left[\left[\bf{3}_\phi\bf{3}_\phi\right]_{\bf{2}}\left[\bf{3}_\theta\bf{3}_\theta\right]_{\bf{2}}\right]_{\bf{1}},\,\,\,\,\,\,\gamma_3\left[\left[\bf{3}_\phi\bf{3}_\phi\right]_{\bf{3}}\left[\bf{3}_\theta\bf{3}_\theta\right]_{\bf{3}}\right]_{\bf{1}}, \,\,\,\,\,\, \text{and} \,\,\,\,\,\, \gamma_4\left[\left[\bf{3}_\phi\bf{3}_\theta\right]_{\bf{2}}\left[\bf{3}_\theta\bf{3}_\phi\right]_{\bf{2}}\right]_{\bf{1}}
\end{equation}
contractions represent a complete and independent basis, where we have labeled these contractions with an associated coupling $\gamma_i$.  We will not write the explicit form of $V_{\theta\phi}^4$ derived from the group product rules here, for brevity.

\subsubsection*{$\langle \theta \rangle \sim (0,0,1)$ and $\langle \phi \rangle \sim (1,1,1)$}
We first \emph{assume} that two of the triplet alignments explored in Section \ref{sec:S4} can be realized simultaneously in \eqref{eq:twoflavonV}, namely $\langle \theta \rangle = v_\theta (0,0,1)$ and $\langle \phi \rangle = v_\phi (1,1,1)$.  Expanding about these points as in \eqref{eq:00vexpand} and \eqref{eq:vvvexpand}, one is led to a potential that represents an expansion about a true minimum of the theory if it respects the RFS associated to said minima, namely
\begin{equation}
\label{eq:RFStwotrip}
\text{RFS} \vert_{\langle \theta \rangle \times \langle \phi \rangle} \sim \left(\mathbb{Z}_2\times \mathbb{Z}_2\right)^\theta \times \text{S3}^{\,\phi}\,,
\end{equation}
where the superscripts indicate the associated flavon VEV.
However, we find that the simultaneous action of the respective RFS generators does not leave $\slashed{V}_{\theta \phi}^4$ invariant,\footnote{In obtaining this result we have iterated through the possible group actions on $\slashed{V}_{\theta\phi}^4$ sourced from all four generators of the RFS, namely $T_{3,3b,123}$ and $G_2$.}
indicating that the RFS is indeed broken by the additional operators.  The symmetry breaking then signals that the \emph{assertion} that we could simultaneously realize the two distinct VEVs exactly as in the single-flavon case was unjustified without further assumptions.  
\subsubsection*{The Emergence of Fine-Tuning}
To prove this, we now go back to \eqref{eq:twoflavonV} and try to identify $\langle \theta_3 \rangle$ and $\langle \phi_{123} \rangle$ as minima of the unbroken theory by performing the standard manipulations. Again treating the fields as real, such that we only have to work in six dimensions, we find that we can only identify $\langle \theta_3 \rangle$ and $\langle \phi_{123} \rangle$ as critical points if we artificially fine-tune\footnote{For the sake of this discussion, we define `fine-tuning' to be an otherwise ad hoc relationship amongst Lagrangian coefficients/couplings, such that the hyperspace defined by all possible domains of Lagrangian couplings is reduced in dimension.} the mixed-operator couplings of \eqref{eq:twoflavonV}.  As an example, if we impose
\begin{equation}
\label{eq:twotripfinetune1}
\gamma_1 \overset{!}{=} \frac{\gamma_2\,v_\theta^2 + 3 m_\phi^2-6 v_\phi^2\left(3\lambda^\phi_1 + 4 \lambda^\phi_3\right)}{3 v_\theta^2}\,,\quad \gamma_3 = -\frac{\gamma_2}{4}\,,\quad \gamma_4 = -\frac{3 \gamma_2}{2}\,,
\end{equation}
 with $v_{\theta}$ a special function of $v_\phi$, $m_{\theta,\phi}$, and Lagrangian couplings ($v_\phi$ can stay arbitrary at this critical point), we find that the RFS is fully restored:  \eqref{eq:twoflavonV} is invariant under \eqref{eq:RFStwotrip} given \eqref{eq:twotripfinetune1}.

Of course, another form of Lagrangian fine-tuning that will resolve the desired critical points is to simply turn the mixed-operators in \eqref{eq:twoflavonV} off,
\begin{equation}
\label{eq:twotripfinetune2}
\gamma_i \rightarrow 0 \,\,\,\,\, \Longrightarrow \,\,\,\,\,\,\,\,\,\,\,\, v_\theta = \frac{\sqrt{3} \, m_\theta}{\sqrt{6 \lambda^\theta_1 + 4 \lambda^\theta_2}}\,, \,\,\,\,\,\,\,\,\,\, v_\phi = \frac{m_\phi}{\sqrt{6 \lambda^\phi_1 + 8 \lambda^\phi_3}}\,,
\end{equation}
where we see that we have re-derived the conditions on the respective VEVs from the single-flavon analyses in \eqref{eq:00vVEVcondition} and \eqref{eq:vvvVEVcondition}.
Indeed, many models simply assume that mixed-couplings are subdominant to single-flavon couplings in order to demonstrate a desired vacuum alignment.  This is a subtle but persistent form of fine-tuning embedded in such classes of non-Abelian flavour models.

In summary, we have shown that the RFS identified for $\langle \theta_3 \rangle$ and $\langle \phi_{123} \rangle$ in Sections \ref{sec:00v} and \ref{sec:vvv} are not respected by the two-flavon potential in \eqref{eq:twoflavonV} when broken and expanded about $\langle \theta_3 \rangle$ and $\langle \phi_{123} \rangle$, indicating that $\langle \theta_3 \rangle$ and $\langle \phi_{123} \rangle$ can not be simultaneously realized as minima of the theory in the absence of arbitrary fine-tuning of Lagrangian parameters.  
\subsubsection*{Vacuum Reorientation}
To explore the impact of the RFS violation identified above, we again parameterize potential correction(s) to the third-family orientation as
\begin{equation}
\label{eq:thetapert}
\langle \theta_3 \rangle \longrightarrow \langle \theta_3 \rangle^\prime \equiv  \left(\delta_{\theta_1}, \delta_{\theta_2}, v_\theta + \delta_{\theta_3} \right)\,\,\,\,\,\text{with}\,\,\,\,\,\,\delta_{\theta_i} \equiv \delta_{\theta_i} \left(\gamma_i\right),
\end{equation}
 where we have indicated that these corrections are, by definition, functions of the mixed-operator couplings $\gamma$ following from the analysis around \eqref{eq:twotripfinetune2}.  

We treat the $\gamma_i$ couplings as small relative to the $\lambda^{\theta,\phi}_i$ terms and therefore consider them as perturbations to the `leading-order' single-flavon system.  Performing a series analysis in $\gamma_i \rightarrow 0$, and replacing $v_{\theta,\phi}$ with their leading-order expressions in \eqref{eq:00vVEVcondition} and \eqref{eq:vvvVEVcondition}, we find
\begin{align}
\nonumber
\delta_{\theta_1} &\simeq \,\,\,\,\frac{m_\phi^2}{m_\theta} \frac{\sqrt{3\lambda^\theta_1 + 2 \lambda^\theta_2}\,\left(6 \gamma_3-\gamma_4\right)}{\sqrt{96}\left(\lambda^\theta_2 - 3 \lambda^\theta_3\right)\left(3\lambda^\phi_1 + 4\lambda^\phi_3\right)}\,,\\ 
\nonumber
\delta_{\theta_2} &\simeq \,\,\,\,\frac{m_\phi^2}{m_\theta} \frac{\sqrt{3\lambda^\theta_1 + 2 \lambda^\theta_2}\,\left(6 \gamma_3-\gamma_4\right)}{\sqrt{216}\left(\lambda^\theta_2 - 2 \lambda^\theta_3\right)\left(3\lambda^\phi_1 + 4\lambda^\phi_3\right)}\,,\\  
\label{eq:twoflavondeltaperts}
\delta_{\theta_3} &\simeq -\frac{m_\phi^2}{m_\theta} \frac{\left(9 \gamma_1 + 2 \gamma_4\right)}{\sqrt{96}\sqrt{3\lambda^\theta_1 + 2 \lambda^\theta_2}\left(3\lambda^\phi_1 + 4\lambda^\phi_3\right)}\,.
\end{align}
Note that, unlike in \ref{eq:dim6VEVcorrections}, the expressions in \eqref{eq:twoflavondeltaperts} still embed some degree of fine-tuning, given that we have arbitrarily taken the $\gamma_i \rightarrow 0$ limit, albeit not as much as simply taking $\delta_{\theta_i} \rightarrow 0$ as is common in flavoured model building.\footnote{One could solve for $\delta_{\theta}^i$ exactly without taking $\gamma_i \rightarrow 0$, or of course numerically, to avoid tedious expressions.}  Expressions analogous to \eqref{eq:twoflavondeltaperts} can be found for the corrections to the $\phi$ VEV as well.

The main takeaway from \eqref{eq:thetapert} and \eqref{eq:twoflavondeltaperts} is that, in order to avoid extreme fine-tuning, RFS-violating VEVs must be corrected in a way that will immediately impact the phenomenological conclusions of a model.  After all, it is straightforward to see how Yukawa-like couplings $\propto \theta_3 \cdot \theta_i$ will yield different fermionic mass and mixing predictions when taking $\langle \theta_3 \rangle^\prime$ instead of $\langle \theta_3 \rangle$.

\subsection{Toy Model Summary and Further Discussion}
\label{sec:S4summary}

Between Sections \ref{sec:00v} - \ref{sec:vvv} we established that, when the vacua of single-flavon, renormalizable potentials are aligned along special directions in flavour space, the broken-phase Lagrangians respect particular RFS, which we identified for three independent VEVs.  Then, between Sections \ref{sec:d6V}-\ref{sec:twotripV}, we added additional operators to the single-flavon toy models, including $d=6$ EFT operators with one flavon, and two-flavon operators at $d=4$.  We found that terms that violate the RFS identified in Sections \ref{sec:00v} - \ref{sec:vvv} lead to reorientations in the VEV, in the absence of ad hoc fine-tuning of Lagrangian parameters.  

More specifically, we found that all of the single-flavon cases are straightforward in the sense that, even at higher orders ($d=6$), the RFS is preserved and therefore the VEV directions are preserved.  On the other hand, the toy models with multiple flavons revealed some subtleties. In such situations, the linear sum of single-flavon potentials automatically exhibits a direct-product group RFS, with RFS constituents associated to their respective single-flavon potential.  The question then became whether operators that mix flavons also respect the direct-product RFS, which depended on the specific nature of the two-flavon contractions.

We first considered a two-flavon toy model with a triplet aligned along the third family direction, and a non-trivial singlet with no orientation in flavour space.   What is perhaps counter-intuitive compared to the results of the toy model with two triplets discussed below, is that the RFS in this scenario is preserved, and the VEV alignment of the triplet is stable in general (without requiring fine-tuning). Indeed, we found that the lone mixed-flavon operator allowed by the S4 symmetry multiplies a contraction of the triplet that transforms trivially with another contraction of the non-trivial singlet that also transform trivially. With respect to each flavon --- and its respective RFS --- the effect of the mixed term was the same as the effect of the respective bi-linears.

On the other hand, to analyze what is perhaps the most typical situation in flavour model building, we also considered a toy model with two triplets aligned along distinct VEV directions, with different associated RFS. In this case the RFS was, generically speaking, violated by operators that mixed the triplets (though not all mixed-triplet operators do so).  We checked this result in different ways: if we insist on preserving the VEV directions, we conclude that we must restrict the terms in the potential, either with some very specific fine-tuning relating the couplings of distinct mixed-operator couplings, or more simply by setting those couplings to zero (which also amounts to fine-tuning).
It is important to note that, upon enforcing this fine-tuning condition on the Lagrangian, we recovered the overarching RFS. Conversely, in general conditions (absent fine-tuning), the VEV directions could not be maintained as extrema/minima, and we calculated the associated change in direction perturbatively.

\section{Realistic Multi-Flavon Models}
\label{sec:generalize}
The toy models in Section \ref{sec:S4} demonstrate our core messages regarding the correspondence between the simultaneous preservation of RFS and the vacuum alignments of non-Abelian flavour models.  However, most realistic models of flavour introduce three or more flavons into the potential, and involve more complicated shaping symmetries than we have enforced above. Furthermore, in SUSY models, vacuum alignment is often achieved via `driving' superfield scalars. These features change the character of the potential, and even mass dimension at which next-to-leading operators appear.

We now look to phenomenologically viable models in the open literature to test our correspondence principle in these more realistic model-building contexts, namely the $\Delta(27)$ Universal Texture Zero model \cite{deMedeirosVarzielas:2017sdv,Bernigaud:2022sgk} in Section \ref{sec:UTZ}, which is based on a similar alignment mechanism as the toy models in Section \ref{sec:S4}, and the renowned A4 Altarelli-Feruglio model \cite{Altarelli:2005yx} in Section \ref{sec:A4AF}, which employs a more elaborate SUSY alignment mechanism. 

\subsection{$\Delta(27)$ Universal Texture Zero Model}
\label{sec:UTZ}
The Universal Texture Zero (UTZ) introduced in \cite{deMedeirosVarzielas:2017sdv} and further explored in \cite{Bernigaud:2022sgk} builds on earlier $\Delta(27)$ models in the literature \cite{deMedeirosVarzielas:2005qg, deMedeirosVarzielas:2006fc,Ma:2006ip}.
It predicts the spectrum of fermionic mass, mixing, and CP-violation parameters in both the quark and lepton sectors, makes predictions for other observables like neutrinoless double $\beta$ decay, is consistent with broken SUSY, and can be embedded into a Grand Unified Theory (GUT); as such all SM fermions are charged equivalently under $\Delta(27)$ as triplets, with $\Delta(27)$ furnishing both a ${\bf{3}}$ and dual  ${\overline{\bf{3}}}$ irreducible representation (and nine singlet representations labeled ${\bf{1}_{r,s}}$) -- cf. \cite{deMedeirosVarzielas:2005qg, deMedeirosVarzielas:2006fc,Ma:2006ip,Luhn:2007uq,Ishimori:2010au,deMedeirosVarzielas:2015amz}.

We repeat the UTZ's scalar sector and symmetry assignments in Table \ref{tab:UTZcharges}.  The triplet flavons $\theta$ and $\theta_{3,23,123,X}$ are all expected to develop special vacuum alignments in flavour space that break $\Delta(27)$. These alignments are given in \eqref{eq:UTZvevs}.  The singlets $S$ and $\Sigma$ play more subtle roles in shaping the Yukawa sector and implementing the GUT breaking, respectively.

\subsubsection{Single-Flavon Potentials and RFS Identification}
\label{sec:UTZsingleflavon}

\begin{table}[t!]
\centering
    \begin{tabular}{|c|c|c|c|c|c|c|c|c|}
    \hline
    {\bf{\text{Field}}} & $h_{5}$ & $\theta_3$ & $\theta_{23}$ & $\theta_{123}$ & $\theta$ & $\theta_X$ &  $\Sigma$ & $S$ \\
    \hline
    $\Delta(27)$ & ${\bf{1}}_{00}$ & $\overline{{\bf{3}}}$ & $\overline{{\bf{3}}}$ & $\overline{{\bf{3}}}$ & $\overline{{\bf{3}}}$ &  {\bf{3}} & ${\bf{1}}_{00}$  & ${\bf{1}}_{00}$ \\
    \hline
    $\mathbb{Z}_N$ & 0 & 0 & -1 & 2 & 0 & x & 2 & -1\\
    \hline
    \end{tabular}
    \caption{The scalar field and flavour symmetry content of the UTZ Model \cite{deMedeirosVarzielas:2017sdv}.}
    \label{tab:UTZcharges}
\end{table}

The renormalizable potential of an arbitrary single triplet flavon was already given in \eqref{eq:UTZsingleflavonV}.\footnote{The original analysis of UTZ vacuum alignment in \cite{deMedeirosVarzielas:2017sdv} did not consider all four couplings $\lambda_{\theta i}$.} 
Noting that a $\Delta(27)$ singlet operator $1_{r,s}$ transforms as $1_{r,s}\rightarrow \omega^{r,s}1_{r,s}$, that the multiplication of two singlets is trivial if the sum of their indices $r^{(')}+s^{(')}\equiv0\,\mathrm{mod}\,3$, and other group properties found in (e.g.) \cite{Ishimori:2010au,deMedeirosVarzielas:2015amz}, we identify the four distinct $\Delta(27)$ contractions in \eqref{eq:UTZsingleflavonV} as 
\begin{align}
\label{eq:UTZd4contractions}
    \lambda_1\,\left[\bf{\bar3 3}\right]_{\bf{1_{0,0}}}\left[\bf{\bar3 3}\right]_{\bf{1_{0,0}}},\quad \lambda_2\,\left[\bf{\bar3 3}\right]_{\bf{1_{1,0}}}\left[\bf{\bar3 3}\right]_{\bf{1_{2,0}}},\quad \lambda_3\,\left[\bf{\bar3 3}\right]_{\bf{1_{1,2}}}\left[\bf{\bar3 3}\right]_{\bf{1_{2,1}}},\,\mathrm{and}\quad \lambda_4\,\left[\left[\bf{3 3}\right]_{\bf{\bar3}}\left[\bf{\bar3 \bar3}\right]_{\bf{3}}\right]_{{\bf{1_{0,0}}}}.
\end{align}
Here we identified the $\lambda_i$ couplings generically, and will label them with superscripts corresponding to the specific flavon coupling later on.  Using \eqref{eq:UTZsingleflavonV}, \eqref{eq:UTZd4contractions} and the (basis-dependent) group product rules in \cite{deMedeirosVarzielas:2015amz}, one can quickly derive an explicit expression for the single-flavon potential.  Minimizing it, one can study its minima to find that $\langle \theta_3 \rangle$ is realized with 
\begin{equation}
    v_3 = \frac{m_{3}}{\sqrt{2\,(\lambda^{\theta_3}_1+\lambda^{\theta_3}_2+\lambda^{\theta_3}_4})},
\end{equation}
having simplified to the real limit with $\theta_{3} = \theta_{3}^\dagger$. 
As in Section \ref{sec:S4}, it is straightforward to identify that 
\begin{equation}
\label{eq:T3UTZ}
T_{3c} \cdot \langle \theta_3 \rangle = \langle \theta_3 \rangle \quad \quad \text{with} \quad \quad T_{3c} = G_a\cdot G_b\cdot G_a\cdot G_a = \mathrm{Diag}(\omega^{-1},\omega,1)\,, 
\end{equation}
where $\omega^3 = 1$ and $G_a$ and $G_b$ are the generators of $\Delta(27)$.  In the triplet basis we work in, they are given by
\begin{equation}
\label{eq:D27gens}
G_a=\begin{pmatrix}
        0&1&0\\
        0&0&1\\
        1&0&0
    \end{pmatrix},\quad
    G_b=\begin{pmatrix}
        1&0&0\\
        0&1/\omega&0\\
        0&0&\omega
    \end{pmatrix}.
\end{equation}
It is clear that the total RFS associated to $\langle \theta_3 \rangle$ is a $\mathbb{Z}_3$.\footnote{For both $\langle \theta_3 \rangle$ and $\langle \theta_{123} \rangle$, there are two non-trivial $\Delta(27)$ elements that leave each VEV invariant.  However, it is easily shown that $T_{3c}$ ($G_a$) generates the remaining non-trivial element and the identity, rounding out the complete $\mathbb{Z}_3$ RFS associated to $\langle \theta_3 \rangle$ ($\langle \theta_{123} \rangle$).}  Similarly, one can deduce (in the real limit) that $\langle \theta_{123} \rangle$ is realized if
\begin{equation}
v_{123} = \frac{\sqrt{3}\,m_{123}}{\sqrt{6\,\lambda^{\theta_{123}}_1+2\lambda^{\theta_{123}}_4}}\,.
\end{equation}
The RFS this alignment respects is even simpler to deduce, as the VEV is left invariant under the action of the group generator $G_a$,
\begin{equation}
\label{eq:GaUTZ}
G_a \cdot \langle \theta_{123} \rangle = \langle \theta_{123} \rangle\,,
\end{equation}
leading to another $\mathbb{Z}_3$ subgroup of $\Delta(27)$.

On the other hand, a close inspection of \eqref{eq:UTZsingleflavonV} reveals that neither $\langle \theta_{23} \rangle$ nor $\langle \theta_X \rangle$ can be realized in single-flavon theories.  Indeed, it was argued in \cite{deMedeirosVarzielas:2017sdv} that $\theta_{X,23}$ are only realized in the $\Delta(27)$ potential when operators $\sim \lbrace \theta_{X,i} \theta_{123}^{\dagger,i} \theta_{123,j} \theta_X^{\dagger,j},\,\, \theta_{23,i} \theta_X^i \theta_{23,j}^\dagger \theta_X^{\dagger,j},\,\, \theta_{23,i} \theta_{3}^{\dagger,i} \theta_{3,j} \theta_{23}^{\dagger,j} \rbrace$, along with a cubic term in $\theta_X$, are present and with specific constraints on their couplings.  That these VEVs cannot even be realized from $\eqref{eq:UTZsingleflavonV}$ indicates that \emph{no} RFS should be associated to them.  Upon using $G_{a,b}$ to generate all 27 triplet representations of the $\Delta(27)$ group elements, we indeed find that there are no elements $G_i$ that satisfy $G_i \cdot \langle \theta_{23,X} \rangle = \langle \theta_{23,X} \rangle$ outside of the identity element $\mathbb{1}$.  This calculation amounts to yet another consistency check on our correspondence principle.  In summary, the RFS we study for the UTZ model is given by
\begin{equation}
\label{eq:RFSUTZ}
\text{RFS} \big \vert_{\text{UTZ}} \sim \mathbb{Z}_3^{\theta_3} \times \mathbb{Z}_3^{\theta_{123}} \times \mathbb{1}^{\theta_{23}} \times \mathbb{1}^{\theta_X}\,.
\end{equation}
Given the lessons of Section \ref{sec:S4}, we should only expect \eqref{eq:RFSUTZ} to hold in the linear combination of single-flavon potentials,
\begin{equation}
\label{eq:V4linearflavons}
V^4_{\text{UTZ}} \supset V^4_{\theta_3} + V^4_{\theta_{123}} + V^4_{\theta_{23}} + V^4_{\theta_{X}}\,.
\end{equation}
We have confirmed this by breaking the $\Delta(27)$ symmetry of \eqref{eq:V4linearflavons} with \eqref{eq:UTZvevs}, and then expanding those VEVs as in (e.g.) \eqref{eq:00vexpand}, \eqref{eq:vvvexpand}, etc.  Transforming the real fields in the resulting expression under the action of $T_{3c}$ and $G_a$, we find that \eqref{eq:V4linearflavons} is indeed invariant under the RFS.

\subsubsection{Mixed-Flavon Operators and RFS Violation}
\label{sec:UTZmultiflavon}
The Lagrangian in \eqref{eq:V4linearflavons} does not yet include operators with multiple flavons.  However, a {\tt{DECO}} analysis reveals that, even when only two of the flavons in Table \ref{tab:UTZcharges} are present, there are over 40 independent operators in the Lagrangian at $d=4$.  Since there are five 
operators in each term of \eqref{eq:V4linearflavons}, that leaves over 30 operators coupling the two flavons together.  
A core lesson of Section \ref{sec:S4} is that mixed operators with distinct triplets is likely to break the RFS.  

We will not calculate all relevant mixed-flavon contractions, but note that they include operators of the form 
\begin{equation}
\label{eq:UTZmixops}
\mathcal{O}_1 \sim \gamma_1\,\left[\bf{\bar3_{\theta_A} 3_{\theta_B}}\right]_{\bf{1}}\left[\bf{\bar3_{\theta_A} 3_{\theta_B}}\right]_{\bf{1}} \quad \text{and} \quad \mathcal{O}_2 \sim \gamma_2\,\left[\left[\bf{\bar3_{\theta_A} \bar3_{\theta_A}}\right]_{\bf{3}}\left[\bf{3_{\theta_B} 3_{\theta_B}}\right]_{\bf{\bar3}}\right]_{{\bf{1}}}\,.
\end{equation}
Taking $\theta_A \equiv \theta_3$ and $\theta_B \equiv \theta_{123}$ as exemplar VEVs, defining $V^4_{\text{UTZ}}$ by adding \eqref{eq:UTZmixops} to \eqref{eq:UTZsingleflavonV}, and breaking the overarching $\Delta(27)$ symmetry via \eqref{eq:00vexpand} and \eqref{eq:vvvexpand}, we find that the scalar RFS in \eqref{eq:RFSUTZ} is \emph{broken}.  This symmetry violation is clearly sourced by the mixed operators of \eqref{eq:UTZmixops}, a fact that is consistent with the mixed-triplet analysis in Section \ref{sec:twotripV} above.  However, the situation is even more subtle for the present $\Delta(27)$-invariant Lagrangian, as a careful analysis shows that 
\begin{equation}
\label{eq:UTZRFSinvariance}
T_{3c} \left[\slashed{V}^4_{\text{UTZ}} \right] = \left[\slashed{V}^4_{\text{UTZ}} \right] \quad \text{while} \quad G_a \left[\slashed{V}^4_{\text{UTZ}} \right] \neq \left[\slashed{V}^4_{\text{UTZ}} \right]\,.
\end{equation}
This implies that the RFS of \eqref{eq:RFSUTZ} is only being broken by the $\mathbb{Z}_3^{\theta_{123}}$ subgroup associated to the $\langle \theta_{123} \rangle$ VEV.\footnote{We are working in the complex limit in this section, given the presence of the dual $\bar3$ representation.}

The correspondence principle developed in Section \ref{sec:S4} tells us that we should expect corrections to the $\langle \theta_{3,123} \rangle$ VEVs as defined in \eqref{eq:VEVcorrect1}.  Performing a perturbative calculation in the $\gamma_{1,2} \rightarrow 0$ limit, we find that 
\begin{align}
\nonumber
&\delta_{\theta_{3_1}} = \delta_{\theta_{3_2}} \simeq 0, \,\,\,\, \delta_{\theta_{3_3}} \simeq - \frac{m_{123}^2 \left(\gamma_1 + \gamma_2 \right)}{\sqrt{32}\,m_3\left(3 \lambda^{\theta_{123}}_1 + \lambda^{\theta_{123}}_4\right)\sqrt{\lambda^{\theta_3}_1 + \lambda^{\theta_3}_2 + \lambda^{\theta_3}_4}} \,,
\\
\label{eq:UTZvevcorrect}
&\delta_{\theta_{123_1}} \neq \delta_{\theta_{123_2}} \neq \delta_{\theta_{123_3}} \neq 0\,,
\end{align}
where we calculated in the real limit, and where the exact expressions for $\delta_{\theta_{123_i}}$ are omitted for brevity.  While these expressions are in no way complete, the manifestation of the RFS correspondence principle is clear; the $\langle \theta_3 \rangle$ VEV is not misaligned as a result of \eqref{eq:UTZmixops}, while the $\langle \theta_{123} \rangle$ VEV \emph{is}.  This result is fully consistent with \eqref{eq:UTZRFSinvariance}!  Of course, there are many more mixed operators beyond \eqref{eq:UTZmixops}, and the expectation is that the RFS will be completely broken by some of them.  We will leave the calculation of complete corrections to the UTZ VEVs to future work, where the phenomenological implications of \eqref{eq:UTZvevcorrect} can be explored more comprehensively.

\subsection{A4 Altarelli-Feruglio Model}
\label{sec:A4AF}
The UTZ of Section \ref{sec:UTZ} largely followed the patterns first observed in the S4 toy models of Section \ref{sec:S4}.  Its alignment mechanism assumed fine-tuned hierarchies between couplings of mixed-flavon operators even at the renormalizable level.  Correspondingly, we found it broke the residual symmetry identified for the constituent flavons, and calculated the corresponding (partial) correction to the VEVs.   

In SUSY contexts, the alignments presented in Sections \ref{sec:S4} and \ref{sec:UTZ} are known as D-term alignments, as used e.g. in \cite{deMedeirosVarzielas:2006fc}.\footnote{Although note that those alignments can proceed without assuming SUSY, as we clearly demonstrated.}  However, many SUSY models implement an F-term alignment mechanism, where specific alignment (or driving) superfields are used, together with an R-symmetry and shaping symmetry, such that the vanishing F-terms with respect to the alignment superfields drive the direction for the scalar components of the flavon superfields.  This includes the renowned Altarelli-Feruglio (AF) model \cite{Altarelli:2005yx}, one of the earliest models to predict the infamous `tri-bimaximal' lepton mixing matrix \cite{Harrison:2002er} at leading order.\footnote{Although \cite{Altarelli:2005yx} demonstrate how this prediction is corrected in a phenomenologically favorable way, upon including higher-order effects.}  AF is based on a broken A4 flavour symmetry, where A4 is the alternating group of 12 elements.  It is isomorphic to $\Delta(12)$, and represents the symmetry group of a tetrahedron.  The AF model mainly describes lepton flavour patterns while predicting the CKM quark mixing matrix to be unity at leading order. Vacuum alignment in the AF model is addressed in great detail between Section 4 and Appendix B of \cite{Altarelli:2005yx}, and we repeat some of that discussion in order to explore our correspondence principle below.

\begin{table}[t!]
\centering
    \begin{tabular}{|c|c|c|c|c|c|c|c|c|}
    \hline
    {\bf{\text{Field}}} & $h_{u,d}$ & $\phi_{T}$ & $\phi_{S}$ & $\xi$ & $\tilde{\xi}$ & $\phi_0^{T}$ &  $\phi_0^{S}$ & $\xi_0$ \\
    \hline
    $A_4$ & {\bf{1}} & {\bf{3}} & {\bf{3}} & {\bf{1}} & {\bf{1}} &  {\bf{3}}& {\bf{3}} & {\bf{1}}\\
    \hline
    $\mathbb{Z}_3$ & {\bf{1}} & {\bf{1}} & {\bf{$\omega$}} & {\bf{$\omega$}} & {\bf{$\omega$}} &  {\bf{1}}& {\bf{$\omega$}} & {\bf{$\omega$}}\\
    \hline
    \end{tabular}
    \caption{The scalar field and flavour symmetry content of the Altarelli-Feruglio Model \cite{Altarelli:2005yx}.}
    \label{tab:AFcharges}
\end{table}
\subsubsection{Leading-Order Driving Potential}
The F-term SUSY alignment mechanism embedded in AF is driven by the minimization of a driving superpotential $w_d$,
\begin{equation}
\label{eq:AFdriving}
    w_d = M \left(\phi_0^T \phi_T \right)+ g \left(\phi_0^T \phi_T \phi_T \right) + g_1 \left(\phi_0^S \phi_S \phi_S \right) + g_2\, \tilde{\xi} \left(\phi_0^S \phi_S \right) + g_3\, \xi_0 \left(\phi_S\phi_S\right) + g_4\,\xi_0 \xi^2 + g_5\, \xi_0 \xi \tilde{\xi} + g_6\, \xi_0 \tilde{\xi}^2\,,
\end{equation}
with respect to `driving' superfields $\lbrace \phi, \xi \rbrace_0$.  Here $\phi_{T,S}$ are the flavons responsible for shaping the Yukawa sector upon reaching their minima and breaking the overarching A4 flavour symmetry of the theory. They are charged as triplets under A4, recalling that A4 is the smallest non-Abelian discrete symmetry that furnishes a three-dimensional irreducible representation.  The driving fields $\phi_{S,T}^0$ are also A4 triplets, while all $\xi_i$ terms are trivial singlets.  The scalar field and symmetry content of the model is reproduced in Table \ref{tab:AFcharges} for completeness, where an additional Abelian $Z_3$ shaping symmetry is present that forbids certain operator couplings, as in our S4 toy models.  There is an additional $U(1)_R$ symmetry present that incorporates R-parity violation into the AF model, but we have omitted presenting it for brevity. 

In the limit of unbroken SUSY, the overall scalar potential of the theory is given by 
\begin{equation}
\label{eq:AFV}
V_{AF}= \sum_i \bigg| \frac{\partial w}{\partial \phi_i} \bigg|^2\,,
\end{equation}
where the derivatives of $w$, a combination of the driving potential $w_d$ and the SUSY Yukawa potential $w_l$ not shown here, is across all scalar fields in the theory.  Minimizing \eqref{eq:AFV}, AF finds that scalar orientations in flavour space corresponding to
\begin{align}
\label{eq:AFvevs}
\langle \phi_T \rangle &= \left(v_T, 0, 0\right)\,,\,\,\,\,\,\,\,\,\,\,\langle \phi_S \rangle = \left(v_S, v_S, v_S\right)\,,\,\,\,\,\,\,\,\,\,\,\langle \xi \rangle = u\,,\,\,\,\,\,\,\,\,\,\,\,\langle \tilde{\xi} \rangle = 0 \,,
\end{align}
where $v_i$ and $u$ are dimensionful, 
\begin{equation}
v_T = -\frac{3 M}{2 g}\,, \,\,\,\,\,\,\,\,\,\, v_S^2 = - \frac{g_4}{3 g_3} u^2 \,,
\end{equation}
and while all driving fields have zero VEV in this limit, correspond to a minimum of the scalar potential that breaks A4.

Critical to our discussion, AF also identifies a potential scalar RFS corresponding to both $\langle \phi_T \rangle$ and $\langle \phi_S \rangle$, such that 
\begin{equation}
    G_T \langle \phi_T \rangle = \langle \phi_T \rangle, \,\,\,\,\,\,\,\,\,\, G_S \langle \phi_S \rangle = \langle \phi_S \rangle\,.
\end{equation}
Here $G_{T,S}$ are representations of the generators of A4, and are given by
\begin{equation}
\label{eq:A4gens}
G_T = \left( 
\begin{array}{ccc}
1 & 0 & 0 \\
0 & \omega^2 & 0 \\
0 & 0 & \omega
\end{array}
\right)\,, \,\,\,\,\,\,\,\,\,\,
G_S = \frac{1}{3}\left(
\begin{array}{ccc}
-1 & 2 & 2 \\
2 & -1 & 2 \\
2 & 2 & -1
\end{array}
\right)\,,
\end{equation}
where $\omega \equiv e^{2\pi i / 3}$.  $G_T$ and $G_S$ respectively generate $\mathbb{Z}_3$ and $\mathbb{Z}_2$ subgroups of A4, such that the overall group `presentation' is given by
\begin{equation}
A4: \,\,\,\,\,\,\,\,S^2 = (ST)^3 = T^3 = \mathbb{1}\,.
\end{equation}

The analysis of Section \ref{sec:S4} tells us that, in order for the alignments in \eqref{eq:AFvevs} to hold without fine-tuning, the scalar potential should respect the VEV's associated RFS,\footnote{We again scanned over all A4 elements to determine that this is the complete RFS --- two (one) non-trivial element(s)  leave $\langle \phi_T \rangle$ ($\langle \phi_S \rangle$) invariant, which are generated by $G_T$ ($G_S$).} 
\begin{equation}
\label{eq:RFSAF}
\text{RFS}\big\vert_{\text{AF}} \sim \mathbb{Z}^T_3 \times \mathbb{Z}^S_2
\end{equation}
with self-explanatory superscripts.

Applying \eqref{eq:AFvevs} to \eqref{eq:AFV} and then expanding about the resulting expression using the AF analogues to \eqref{eq:00vexpand}, \eqref{eq:vvvexpand}, and the equivalent for the A4 singlets, we find that 
\begin{equation}
\left(G_T \times G_S\right) \left[ \slashed{V}_{AF} \right] = \slashed{V}_{AF}\,,
\end{equation}
indicating that \eqref{eq:RFSAF} represents a true RFS of the broken leading-order model, and therefore that \eqref{eq:AFvevs} represents a true minimum.  One now sees how the SUSY embedded in \eqref{eq:AFV} serves to protect the model against certain fine-tunings even in the flavour sector. 

\subsubsection{Higher-Order Operators and RFS Violation}
\label{sec:AFHO}
In one of the most complete analyses of its kind, Altarelli and Feruglio also examined the corrections to their model coming from higher-order EFT operators in the scalar sector (cf. Appendix B of \cite{Altarelli:2005yx}).  They enumerated a basis of 26 operators invariant under the model's complete symmetry group that augments the leading-order superpotential $w_d$,
\begin{equation}
w_d\,\,\,\,\, \longrightarrow \,\,\,\,\, w_d + \Delta w_d , \,\,\,\,\,\,\,\,\,\,\,\, \Delta w_d = \frac{1}{\Lambda} \left(\sum_{k=3}^{13} t_k \, I^T_k + \sum_{k=1}^{12} s_k \, I_k^S + \sum_{k=1}^{3} x_k\,I_k^X \right)],
\end{equation}
where the superscripts on the operator classes $I_k^{T,S,X}$ indicate the driving field associated with the sector.  For example, 
\begin{equation}
\label{eq:sampleHOAF}
I_3^T = \left[\phi_0^T \phi_T\right]_{{\bf{1}}}\left[\phi_T\phi_T\right]_{{\bf{1}}},\,\,\,\,\,I_7^S = \left[\phi_0^S \left[\phi_T\phi_S\right]_{{\bf{A}}}\right]_{{\bf{1}}} \tilde{\xi},\,\,\,\,I_1^X =\xi_0 \left[\phi_T \left[\phi_T\phi_S\right]_{{\bf{S}}}\right]_{{\bf{1}}}\,,
\end{equation}
with brackets corresponding to A4 group contractions, and the subscripts ${{\bf{S, (A)}}}$ indicating a contraction to an (anti)symmetric triplet.\footnote{We found a typo (a missing minus sign) in the third component of the RHS of equation (9) in \cite{Altarelli:2005yx}, where the antisymmetric triplet contraction is defined. That is, we find that the contraction between two triplets $a$ and $b$ should transform as $\left[ab\right]_{{\bf{A}}} = \frac{1}{2} \left(a_2 b_3 - a_3 b_2, a_1 b_2 - a_2 b_1, a_3 b_1 - a_1 b_3 \right)$.}

We now ask whether the operators of $\Delta w_d$ also respect \eqref{eq:RFSAF} in the broken phase. Looking to \eqref{eq:sampleHOAF} for hints, we observe that the potential includes a mixed $\phi_S-\phi_T$ contraction in $I_7^S$ and $I_1^X$, and a closer look at the complete basis in \cite{Altarelli:2005yx} indicates that only $I_{3,4,5}^T$\footnote{We found another typo in eq. (74) of \cite{Altarelli:2005yx}:  one $\phi_S$ should be relabeled to $\phi_T$ in $I_{2,3}^X$ (e.g. $\left(\phi_S  \phi_S\right) \rightarrow \left(\phi_S  \phi_T\right)$, otherwise these operators are not invariant under the AF $\mathbb{Z}_3$ shaping symmetry.} carry exclusively $S$ or $T$ labels.  We therefore suspect an RFS violation in this case, and correspondingly a vacuum reorientation.  Expanding about the leading-order minimum, we confirm that 
\begin{equation}
\left(G_T \times G_S \right)\left[\slashed{V}_{AF}\right] \neq \slashed{V}_{AF}.
\end{equation}
It is also straightforward to confirm that the unmixed operators $I_{3,4,5}^T$ \emph{do} preserve the RFS, as expected.  

Given the broken RFS, one should expect non-zero perturbations away from the leading order vacuum structure.  Altarelli-Feruglio calculated the corrections defined by
\begin{equation}
\langle \phi_T \rangle \rightarrow \left(v_T + \delta v_1^T, \delta v_2^T, \delta v_3^T \right),\,\,\,\,\,\, \langle \phi_S \rangle \rightarrow v_S + \left(\delta v_1, \delta v_2, \delta v_3 \right),\,\,\,\,\,\,\langle \tilde{\xi} \rangle \rightarrow \delta u^\prime
\end{equation}
sourced from  the entire EFT basis, finding non-zero values for each $\delta$ correction.  We have checked these expressions in eq (76) of \cite{Altarelli:2005yx}, finding complete agreement after correcting for the antisymmetric contraction and the mixed-operator content typos pointed out in Footnotes 17 and 18 below.

\section{Summary and Outlook}
\label{sec:CONCLUDE}
We identified a correspondence between the preservation of a residual flavour symmetry (RFS) in the broken phase of flavon-based models of fermionic mass, mixing, and CP violation, and the stability of special flavour alignments in said flavons' vacuum expectation values (VEV); scalar operators that preserve the RFS allow for stable minima up to generation-independent rescalings, while operators that violate the RFS can reorient the vacua away from preferred, leading-order directions, in the absence of ad hoc fine-tunings of operator couplings. 

We demonstrated this correspondence principle in toy scenarios and realistic models from the literature, which were based on one or many flavons, and built on  three different non-Abelian discrete flavour symmetries (S4, A4, and $\Delta(27)$).  Our findings are therefore substantially model independent, and reveal that (unsurprisingly) operators where multiple flavon multiplets are assumed to develop independent VEVs frequently violate the scalar RFS.  Models like the Altarelli-Feruglio A4 model \cite{Altarelli:2005yx} of Section \ref{sec:A4AF}, which implement an F-term SUSY alignment mechanism, are capable of protecting against this effect at the renormalizable level.  However, we showed that once higher-order EFT operators are present in the scalar potential, even the Altarelli-Feruglio model exhibits scalar RFS violation, and therefore corrections to leading-order VEVs.  This is fully consistent with the relevant calculations presented in \cite{Altarelli:2005yx}.

VEV corrections coming from RFS-violating operators are commonly ignored in the model-building literature, representing an oversight in many predictions on the market.  Following \cite{Altarelli:2005yx}, we also exercised a perturbative method to analytically calculate the VEV corrections in a manner that softens fine-tuning. This directly translates to analytic corrections to the Yukawa sector's predictions.  Of course, the VEV realignments we elaborated in this paper are not the only corrections that can impact the vacuum structure of flavour theories.  For example, quantum corrections to the scalar potential may also lead to realignments.

Another useful way to summarize our results is to note that the vacuum-alignment problem may be turned into a systematic test of the scalar operator basis. Given a desired flavon vacuum structure, one can identify which operators are (in)compatible with it based on RFS. This is not a `solution' to the vacuum alignment problem as in (e.g.) \cite{Babu:2010ex,Holthausen:2011pd}, but rather a means of understanding why those papers are successful at solving it from an underlying RFS perspective.  Alternatively, if one chooses to maintain a flavour symmetry that does \emph{not} preserve a given, leading-order vacuum alignment, our correspondence principle simultaneously demonstrates how to identify the source of any misalignment via RFS-violating operators, and therefore what VEV corrections must be calculated for internal model-building consistency.
Our framework therefore provides a practical route from vacuum alignment as an input assumption to vacuum alignment as a controlled property of the underlying theory.

Our work leads to further questions of interest.  After all, it is  clear that a complete numerical analysis of VEV corrections in models like the Universal Texture Zero \cite{deMedeirosVarzielas:2017sdv,Bernigaud:2022sgk} studied in Section \ref{sec:UTZ} may be required for full internal consistency of its predictions. Alternatively, a UTZ model might be constructed using an F-term SUSY alignment mechanism following the prescription in \cite{deMedeirosVarzielas:2015amz}.  Perhaps the largest question raised by our analysis has to do with the foundational choice of field and symmetry content of a flavour model: can corrections to VEV alignments require fewer flavons, or lower orders in Yukawa sector EFT expansions, to realize hierarchical textures in fermionic mass matrices?    We leave these and other questions to future work.

\section*{Acknowledgements}
We thank Rudi Rahn for inspiring and helpful conversations on the topic, including detailed support as we augmented the {\tt{DECO}} package to account for $\Delta(27)$. 
IdMV thanks the University of Basel for hospitality.
IdMV acknowledges funding from Fundação para a Ciência e a Tecnologia (FCT) through the FCT Mobility program, and through
the projects CFTP-FCT Unit 
UID/00777/2025 (\url{https://doi.org/10.54499/UID/00777/2025}),  UIDB/FIS/00777/2020 and UIDP/FIS/00777/2020, CERN/FIS-PAR/0019/2021,
CERN/FIS-PAR/0002/2021, 2024.02004 CERN, which are partially funded through POCTI (FEDER), COMPETE,
QREN and EU.
M.-S. L. is thankful for the Center for Computational Astrophysics, Flatiron Institute for hosting him as a Guest Researcher. 
AS is partially supported by the U.S. National
Science Foundation, under the Grant No. PHY-2310363 and also by NSF, under the Grant No. NSF OAC-2417682.
JT is especially grateful to LANL's Theoretical Division, and Christopher Lee in particular, for hosting his Guest Researcher Agreement.  LANL is operated by Triad National Security, LLC, for the National Nuclear Security Administration of U.S. Department of Energy (Contract No. 89233218CNA000001).

\bibliographystyle{unsrt}
\bibliography{bibliovacuumRFS}

@article{deMedeirosVarzielas:2005qg,
    author = "de Medeiros Varzielas, I. and King, S. F. and Ross, G. G.",
    title = "{Tri-bimaximal neutrino mixing from discrete subgroups of SU(3) and SO(3) family symmetry}",
    eprint = "hep-ph/0512313",
    archivePrefix = "arXiv",
    doi = "10.1016/j.physletb.2006.11.015",
    journal = "Phys. Lett. B",
    volume = "644",
    pages = "153--157",
    year = "2007"
}

@article{deMedeirosVarzielas:2021zqs,
    author = "de Medeiros Varzielas, Ivo and Ivanov, Igor P. and Levy, Miguel",
    title = "{Exploring multi-Higgs models with softly broken large discrete symmetry groups}",
    eprint = "2107.08227",
    archivePrefix = "arXiv",
    primaryClass = "hep-ph",
    doi = "10.1140/epjc/s10052-021-09681-w",
    journal = "Eur. Phys. J. C",
    volume = "81",
    number = "10",
    pages = "918",
    year = "2021"
}

@article{King:2009ap,
    author = "King, Stephen F. and Luhn, Christoph",
    title = "{On the origin of neutrino flavour symmetry}",
    eprint = "0908.1897",
    archivePrefix = "arXiv",
    primaryClass = "hep-ph",
    doi = "10.1088/1126-6708/2009/10/093",
    journal = "JHEP",
    volume = "10",
    pages = "093",
    year = "2009"
}

@article{Hagedorn:2023mrg,
    author = "Hagedorn, Claudia and Lopez-Ibanez, M. L. and Perez, M. Jay and Rahat, Moinul Hossain and Vives, Oscar",
    title = "{Flavon vacuum alignment beyond SUSY}",
    eprint = "2312.07430",
    archivePrefix = "arXiv",
    primaryClass = "hep-ph",
    reportNumber = "IFIC/23-25, FTUV/23-0718",
    doi = "10.1103/PhysRevD.110.015009",
    journal = "Phys. Rev. D",
    volume = "110",
    number = "1",
    pages = "015009",
    year = "2024"
}

@article{Lam:2007qc,
    author = "Lam, C. S.",
    title = "{Symmetry of Lepton Mixing}",
    eprint = "0708.3665",
    archivePrefix = "arXiv",
    primaryClass = "hep-ph",
    doi = "10.1016/j.physletb.2007.09.032",
    journal = "Phys. Lett. B",
    volume = "656",
    pages = "193--198",
    year = "2007"
}

@article{Hernandez:2012ra,
    author = "Hernandez, D. and Smirnov, A. Yu.",
    title = "{Lepton mixing and discrete symmetries}",
    eprint = "1204.0445",
    archivePrefix = "arXiv",
    primaryClass = "hep-ph",
    doi = "10.1103/PhysRevD.86.053014",
    journal = "Phys. Rev. D",
    volume = "86",
    pages = "053014",
    year = "2012"
}

@article{Lam:2012ga,
    author = "Lam, C. S.",
    title = "{Finite Symmetry of Leptonic Mass Matrices}",
    eprint = "1208.5527",
    archivePrefix = "arXiv",
    primaryClass = "hep-ph",
    doi = "10.1103/PhysRevD.87.013001",
    journal = "Phys. Rev. D",
    volume = "87",
    number = "1",
    pages = "013001",
    year = "2013"
}

@article{Holthausen:2012wt,
    author = "Holthausen, Martin and Lim, Kher Sham and Lindner, Manfred",
    title = "{Lepton Mixing Patterns from a Scan of Finite Discrete Groups}",
    eprint = "1212.2411",
    archivePrefix = "arXiv",
    primaryClass = "hep-ph",
    doi = "10.1016/j.physletb.2013.02.047",
    journal = "Phys. Lett. B",
    volume = "721",
    pages = "61--67",
    year = "2013"
}

@article{King:2013vna,
    author = "King, Stephen F. and Neder, Thomas and Stuart, Alexander J.",
    title = "{Lepton mixing predictions from $\Delta(6n^2)$ family Symmetry}",
    eprint = "1305.3200",
    archivePrefix = "arXiv",
    primaryClass = "hep-ph",
    doi = "10.1016/j.physletb.2013.08.052",
    journal = "Phys. Lett. B",
    volume = "726",
    pages = "312--315",
    year = "2013"
}

@article{Holthausen:2013vba,
    author = "Holthausen, Martin and Lim, Kher Sham",
    title = "{Quark and Leptonic Mixing Patterns from the Breakdown of a Common Discrete Flavor Symmetry}",
    eprint = "1306.4356",
    archivePrefix = "arXiv",
    primaryClass = "hep-ph",
    doi = "10.1103/PhysRevD.88.033018",
    journal = "Phys. Rev. D",
    volume = "88",
    pages = "033018",
    year = "2013"
}

@article{Lavoura:2014kwa,
    author = "Lavoura, Luis and Ludl, Patrick Otto",
    title = "{Residual $\mathbb{Z}_2 \times \mathbb{Z}_2$ symmetries and lepton mixing}",
    eprint = "1401.5036",
    archivePrefix = "arXiv",
    primaryClass = "hep-ph",
    reportNumber = "CFTP-14-002, UWTHPH-2014-5",
    doi = "10.1016/j.physletb.2014.03.001",
    journal = "Phys. Lett. B",
    volume = "731",
    pages = "331--336",
    year = "2014"
}

@article{Fonseca:2014koa,
    author = "Fonseca, Renato M. and Grimus, Walter",
    title = "{Classification of lepton mixing matrices from finite residual symmetries}",
    eprint = "1405.3678",
    archivePrefix = "arXiv",
    primaryClass = "hep-ph",
    reportNumber = "UWTHPH-2014-11, IFIC-14-32",
    doi = "10.1007/JHEP09(2014)033",
    journal = "JHEP",
    volume = "09",
    pages = "033",
    year = "2014"
}

@article{Joshipura:2014qaa,
    author = "Joshipura, Anjan S. and Patel, Ketan M.",
    title = "{Discrete flavor symmetries for degenerate solar neutrino pair and their predictions}",
    eprint = "1405.6106",
    archivePrefix = "arXiv",
    primaryClass = "hep-ph",
    doi = "10.1103/PhysRevD.90.036005",
    journal = "Phys. Rev. D",
    volume = "90",
    number = "3",
    pages = "036005",
    year = "2014"
}

@article{Talbert:2014bda,
    author = "Talbert, Jim",
    title = "{[Re]constructing Finite Flavour Groups: Horizontal Symmetry Scans from the Bottom-Up}",
    eprint = "1409.7310",
    archivePrefix = "arXiv",
    primaryClass = "hep-ph",
    doi = "10.1007/JHEP12(2014)058",
    journal = "JHEP",
    volume = "12",
    pages = "058",
    year = "2014"
}

@article{Yao:2015dwa,
    author = "Yao, Chang-Yuan and Ding, Gui-Jun",
    title = "{Lepton and Quark Mixing Patterns from Finite Flavor Symmetries}",
    eprint = "1505.03798",
    archivePrefix = "arXiv",
    primaryClass = "hep-ph",
    doi = "10.1103/PhysRevD.92.096010",
    journal = "Phys. Rev. D",
    volume = "92",
    number = "9",
    pages = "096010",
    year = "2015"
}

@article{Lu:2016jit,
    author = "Lu, Jun-Nan and Ding, Gui-Jun",
    title = "{Alternative Schemes of Predicting Lepton Mixing Parameters from Discrete Flavor and CP Symmetry}",
    eprint = "1610.05682",
    archivePrefix = "arXiv",
    primaryClass = "hep-ph",
    reportNumber = "USTC-ICTS-16-11",
    doi = "10.1103/PhysRevD.95.015012",
    journal = "Phys. Rev. D",
    volume = "95",
    number = "1",
    pages = "015012",
    year = "2017"
}

@article{deMedeirosVarzielas:2016fqq,
    author = "de Medeiros Varzielas, Ivo and Rasmussen, Rasmus W. and Talbert, Jim",
    title = "{Bottom-Up Discrete Symmetries for Cabibbo Mixing}",
    eprint = "1605.03581",
    archivePrefix = "arXiv",
    primaryClass = "hep-ph",
    reportNumber = "DESY-16-125",
    doi = "10.1142/S0217751X17500476",
    journal = "Int. J. Mod. Phys. A",
    volume = "32",
    number = "06n07",
    pages = "1750047",
    year = "2017"
}

@article{deMedeirosVarzielas:2019lgb,
    author = "de Medeiros Varzielas, Ivo and Talbert, Jim",
    title = "{Simplified Models of Flavourful Leptoquarks}",
    eprint = "1901.10484",
    archivePrefix = "arXiv",
    primaryClass = "hep-ph",
    reportNumber = "DESY-18-210, DESY 18-210",
    doi = "10.1140/epjc/s10052-019-7047-2",
    journal = "Eur. Phys. J. C",
    volume = "79",
    number = "6",
    pages = "536",
    year = "2019"
}

@article{Varzielas:2023qlb,
    author = "Varzielas, Ivo de Medeiros and Sengupta, Amartya",
    title = "{Constraining flavoured leptoquarks with LHC and LFV}",
    eprint = "2301.04119",
    archivePrefix = "arXiv",
    primaryClass = "hep-ph",
    doi = "10.1016/j.nuclphysb.2024.116495",
    journal = "Nucl. Phys. B",
    volume = "1001",
    pages = "116495",
    year = "2024"
}

@article{Desai:2023jxh,
    author = "Desai, Nishita and Sengupta, Amartya",
    title = "{Status of leptoquark models after LHC Run-2 and discovery prospects at future colliders}",
    eprint = "2301.01754",
    archivePrefix = "arXiv",
    primaryClass = "hep-ph",
    month = "1",
    year = "2023"
}

@article{Bernigaud:2019bfy,
    author = "Bernigaud, Jordan and de Medeiros Varzielas, Ivo and Talbert, Jim",
    title = "{Finite Family Groups for Fermionic and Leptoquark Mixing Patterns}",
    eprint = "1906.11270",
    archivePrefix = "arXiv",
    primaryClass = "hep-ph",
    reportNumber = "LAPTH-033/19, DESY-19-091, DESY 19-091",
    doi = "10.1007/JHEP01(2020)194",
    journal = "JHEP",
    volume = "01",
    pages = "194",
    year = "2020"
}

@article{Bernigaud:2020wvn,
    author = "Bernigaud, Jordan and de Medeiros Varzielas, Ivo and Talbert, Jim",
    title = "{Reconstructing Effective Lagrangians Embedding Residual Family Symmetries}",
    eprint = "2005.12293",
    archivePrefix = "arXiv",
    primaryClass = "hep-ph",
    reportNumber = "P3H-20-021, TTP20-022",
    doi = "10.1140/epjc/s10052-021-08882-7",
    journal = "Eur. Phys. J. C",
    volume = "81",
    number = "1",
    pages = "65",
    year = "2021"
}

@article{deMedeirosVarzielas:2019dyu,
    author = "de Medeiros Varzielas, Ivo and Talbert, Jim",
    title = "{FCNC-free multi-Higgs-doublet models from broken family symmetries}",
    eprint = "1908.10979",
    archivePrefix = "arXiv",
    primaryClass = "hep-ph",
    reportNumber = "DESY 19-147, DESY-19-147",
    doi = "10.1016/j.physletb.2019.135091",
    journal = "Phys. Lett. B",
    volume = "800",
    pages = "135091",
    year = "2020"
}

@article{Harrison:2002er,
    author = "Harrison, P. F. and Perkins, D. H. and Scott, W. G.",
    title = "{Tri-bimaximal mixing and the neutrino oscillation data}",
    eprint = "hep-ph/0202074",
    archivePrefix = "arXiv",
    reportNumber = "RAL-TR-2002-002",
    doi = "10.1016/S0370-2693(02)01336-9",
    journal = "Phys. Lett. B",
    volume = "530",
    pages = "167",
    year = "2002"
}

@article{deMedeirosVarzielas:2015amz,
    author = "de Medeiros Varzielas, Ivo",
    title = "{$\Delta(27)$ family symmetry and neutrino mixing}",
    eprint = "1507.00338",
    archivePrefix = "arXiv",
    primaryClass = "hep-ph",
    doi = "10.1007/JHEP08(2015)157",
    journal = "JHEP",
    volume = "08",
    pages = "157",
    year = "2015"
}

@article{Altarelli:2010gt,
    author = "Altarelli, Guido and Feruglio, Ferruccio",
    title = "{Discrete Flavor Symmetries and Models of Neutrino Mixing}",
    eprint = "1002.0211",
    archivePrefix = "arXiv",
    primaryClass = "hep-ph",
    reportNumber = "RM3-TH-10-01, CERN-PH-TH-2010-016, DFPD-10-TH-02",
    doi = "10.1103/RevModPhys.82.2701",
    journal = "Rev. Mod. Phys.",
    volume = "82",
    pages = "2701--2729",
    year = "2010"
}

@article{Ishimori:2010au,
    author = "Ishimori, Hajime and Kobayashi, Tatsuo and Ohki, Hiroshi and Shimizu, Yusuke and Okada, Hiroshi and Tanimoto, Morimitsu",
    title = "{Non-Abelian Discrete Symmetries in Particle Physics}",
    eprint = "1003.3552",
    archivePrefix = "arXiv",
    primaryClass = "hep-th",
    reportNumber = "KUNS-2260",
    doi = "10.1143/PTPS.183.1",
    journal = "Prog. Theor. Phys. Suppl.",
    volume = "183",
    pages = "1--163",
    year = "2010"
}

@article{King:2013eh,
    author = "King, Stephen F. and Luhn, Christoph",
    title = "{Neutrino Mass and Mixing with Discrete Symmetry}",
    eprint = "1301.1340",
    archivePrefix = "arXiv",
    primaryClass = "hep-ph",
    reportNumber = "IPPP-12-100, DCPT-12-200",
    doi = "10.1088/0034-4885/76/5/056201",
    journal = "Rept. Prog. Phys.",
    volume = "76",
    pages = "056201",
    year = "2013"
}

@article{Alonso:2011yg,
    author = "Alonso, R. and Gavela, M. B. and Merlo, L. and Rigolin, S.",
    title = "{On the scalar potential of minimal flavour violation}",
    eprint = "1103.2915",
    archivePrefix = "arXiv",
    primaryClass = "hep-ph",
    reportNumber = "FTUAM-11-39, IFT-UAM-CSIC-11-09, TUM-HEP-796-11, DFPD-11-TH-2",
    doi = "10.1007/JHEP07(2011)012",
    journal = "JHEP",
    volume = "07",
    pages = "012",
    year = "2011"
}

@article{deMedeirosVarzielas:2006fc,
    author = "de Medeiros Varzielas, I. and King, S. F. and Ross, G. G.",
    title = "{Neutrino tri-bi-maximal mixing from a non-Abelian discrete family symmetry}",
    eprint = "hep-ph/0607045",
    archivePrefix = "arXiv",
    reportNumber = "OUTP-0614P, CERN-PH-TH-2006-124",
    doi = "10.1016/j.physletb.2007.03.009",
    journal = "Phys. Lett. B",
    volume = "648",
    pages = "201--206",
    year = "2007"
}

@article{Ma:2006ip,
    author = "Ma, Ernest",
    title = "{Neutrino Mass Matrix from Delta(27) Symmetry}",
    eprint = "hep-ph/0607056",
    archivePrefix = "arXiv",
    reportNumber = "UCRHEP-T416",
    doi = "10.1142/S0217732306021190",
    journal = "Mod. Phys. Lett. A",
    volume = "21",
    pages = "1917--1921",
    year = "2006"
}

@article{Luhn:2007uq,
    author = "Luhn, Christoph and Nasri, Salah and Ramond, Pierre",
    title = "{The Flavor group Delta(3n**2)}",
    eprint = "hep-th/0701188",
    archivePrefix = "arXiv",
    reportNumber = "UFIFT-HET-07-2",
    doi = "10.1063/1.2734865",
    journal = "J. Math. Phys.",
    volume = "48",
    pages = "073501",
    year = "2007"
}

@article{Loisa:2024xuk,
    author = "Loisa, Eetu and Talbert, Jim",
    title = "{Froggatt-Nielsen meets the SMEFT}",
    eprint = "2402.16940",
    archivePrefix = "arXiv",
    primaryClass = "hep-ph",
    reportNumber = "LA-UR-24-21703",
    doi = "10.1007/JHEP10(2024)017",
    journal = "JHEP",
    volume = "10",
    pages = "017",
    year = "2024"
}

@article{DAmbrosio:2002vsn,
    author = "D'Ambrosio, G. and Giudice, G. F. and Isidori, G. and Strumia, A.",
    title = "{Minimal flavor violation: An Effective field theory approach}",
    eprint = "hep-ph/0207036",
    archivePrefix = "arXiv",
    reportNumber = "CERN-TH-2002-147, IFUP-TH-2002-17",
    doi = "10.1016/S0550-3213(02)00836-2",
    journal = "Nucl. Phys. B",
    volume = "645",
    pages = "155--187",
    year = "2002"
}

@article{Altarelli:2005yp,
    author = "Altarelli, Guido and Feruglio, Ferruccio",
    title = "{Tri-bimaximal neutrino mixing from discrete symmetry in extra dimensions}",
    eprint = "hep-ph/0504165",
    archivePrefix = "arXiv",
    reportNumber = "DFPD-05-TH-14, CERN-PH-TH-2005-067",
    doi = "10.1016/j.nuclphysb.2005.05.005",
    journal = "Nucl. Phys. B",
    volume = "720",
    pages = "64--88",
    year = "2005"
}

@article{Reig:2018ocz,
    author = "Reig, Mario and Valle, Jos\'e W. F. and Wilczek, Frank",
    title = "{SO(3) family symmetry and axions}",
    eprint = "1805.08048",
    archivePrefix = "arXiv",
    primaryClass = "hep-ph",
    reportNumber = "IFIC-XXX, MIT-CTP/5003, MIT-CTP-5003",
    doi = "10.1103/PhysRevD.98.095008",
    journal = "Phys. Rev. D",
    volume = "98",
    number = "9",
    pages = "095008",
    year = "2018"
}

@article{Babu:2002dz,
    author = "Babu, K. S. and Ma, Ernest and Valle, J. W. F.",
    title = "{Underlying A(4) symmetry for the neutrino mass matrix and the quark mixing matrix}",
    eprint = "hep-ph/0206292",
    archivePrefix = "arXiv",
    reportNumber = "UCRHEP-T341, OSU-HEP-02-07, IFIC-02-26",
    doi = "10.1016/S0370-2693(02)03153-2",
    journal = "Phys. Lett. B",
    volume = "552",
    pages = "207--213",
    year = "2003"
}

@article{Altarelli:2005yx,
    author = "Altarelli, Guido and Feruglio, Ferruccio",
    title = "{Tri-bimaximal neutrino mixing, A(4) and the modular symmetry}",
    eprint = "hep-ph/0512103",
    archivePrefix = "arXiv",
    reportNumber = "CERN-PH-TH-2005-226",
    doi = "10.1016/j.nuclphysb.2006.02.015",
    journal = "Nucl. Phys. B",
    volume = "741",
    pages = "215--235",
    year = "2006"
}

@article{King:2001uz,
    author = "King, S. F. and Ross, Graham G.",
    title = "{Fermion masses and mixing angles from SU(3) family symmetry}",
    eprint = "hep-ph/0108112",
    archivePrefix = "arXiv",
    reportNumber = "SHEP-01-21, OUTP-01-46P",
    doi = "10.1016/S0370-2693(01)01139-X",
    journal = "Phys. Lett. B",
    volume = "520",
    pages = "243--253",
    year = "2001"
}

@article{King:2003rf,
    author = "King, S. F. and Ross, Graham G.",
    title = "{Fermion masses and mixing angles from SU (3) family symmetry and unification}",
    eprint = "hep-ph/0307190",
    archivePrefix = "arXiv",
    reportNumber = "SHEP-03-14, CERN-TH-2003-147",
    doi = "10.1016/j.physletb.2003.09.027",
    journal = "Phys. Lett. B",
    volume = "574",
    pages = "239--252",
    year = "2003"
}

@article{deMedeirosVarzielas:2012ylr,
    author = "de Medeiros Varzielas, Ivo and Emmanuel-Costa, David and Leser, Philipp",
    title = "{Geometrical CP Violation from Non-Renormalisable Scalar Potentials}",
    eprint = "1204.3633",
    archivePrefix = "arXiv",
    primaryClass = "hep-ph",
    reportNumber = "CFTP-12-005, DO-TH-12-12",
    doi = "10.1016/j.physletb.2012.08.008",
    journal = "Phys. Lett. B",
    volume = "716",
    pages = "193--196",
    year = "2012"
}

@article{deMedeirosVarzielas:2005ax,
    author = "de Medeiros Varzielas, Ivo and Ross, Graham G.",
    title = "{SU(3) family symmetry and neutrino bi-tri-maximal mixing}",
    eprint = "hep-ph/0507176",
    archivePrefix = "arXiv",
    reportNumber = "OUTP-0420P",
    doi = "10.1016/j.nuclphysb.2005.10.039",
    journal = "Nucl. Phys. B",
    volume = "733",
    pages = "31--47",
    year = "2006"
}

@article{deMedeirosVarzielas:2017sdv,
    author = "de Medeiros Varzielas, Ivo and Ross, Graham G. and Talbert, Jim",
    title = "{A Unified Model of Quarks and Leptons with a Universal Texture Zero}",
    eprint = "1710.01741",
    archivePrefix = "arXiv",
    primaryClass = "hep-ph",
    reportNumber = "OUTP-17-13P, DESY-17-146",
    doi = "10.1007/JHEP03(2018)007",
    journal = "JHEP",
    volume = "03",
    pages = "007",
    year = "2018"
}

@article{Bernigaud:2022sgk,
    author = "Bernigaud, Jordan and Varzielas, Ivo de Medeiros and Levy, Miguel and Talbert, Jim",
    title = "{Revisiting the universal texture zero of flavour: a Markov chain Monte Carlo analysis}",
    eprint = "2211.15700",
    archivePrefix = "arXiv",
    primaryClass = "hep-ph",
    doi = "10.1140/epjc/s10052-023-11654-0",
    journal = "Eur. Phys. J. C",
    volume = "83",
    number = "6",
    pages = "479",
    year = "2023"
}

@article{Alonso:2013hga,
    author = "Alonso, Rodrigo and Jenkins, Elizabeth E. and Manohar, Aneesh V. and Trott, Michael",
    title = "{Renormalization Group Evolution of the Standard Model Dimension Six Operators III: Gauge Coupling Dependence and Phenomenology}",
    eprint = "1312.2014",
    archivePrefix = "arXiv",
    primaryClass = "hep-ph",
    reportNumber = "CERN-PH-TH-2013-305, CERN-PH-TH/2013-305",
    doi = "10.1007/JHEP04(2014)159",
    journal = "JHEP",
    volume = "04",
    pages = "159",
    year = "2014"
}

@article{Buchmuller:1985jz,
    author = "Buchmuller, W. and Wyler, D.",
    title = "{Effective Lagrangian Analysis of New Interactions and Flavor Conservation}",
    reportNumber = "CERN-TH-4254/85",
    doi = "10.1016/0550-3213(86)90262-2",
    journal = "Nucl. Phys. B",
    volume = "268",
    pages = "621--653",
    year = "1986"
}

@article{Grzadkowski:2010es,
    author = "Grzadkowski, B. and Iskrzynski, M. and Misiak, M. and Rosiek, J.",
    title = "{Dimension-Six Terms in the Standard Model Lagrangian}",
    eprint = "1008.4884",
    archivePrefix = "arXiv",
    primaryClass = "hep-ph",
    reportNumber = "IFT-9-2010, TTP10-35",
    doi = "10.1007/JHEP10(2010)085",
    journal = "JHEP",
    volume = "10",
    pages = "085",
    year = "2010"
}

@article{Froggatt:1978nt,
    author = "Froggatt, C. D. and Nielsen, Holger Bech",
    title = "{Hierarchy of Quark Masses, Cabibbo Angles and CP Violation}",
    reportNumber = "CERN-TH-2519",
    doi = "10.1016/0550-3213(79)90316-X",
    journal = "Nucl. Phys. B",
    volume = "147",
    pages = "277--298",
    year = "1979"
}

@article{Rahn:2023deco,
   title={Discrete symmetries and efficient counting of operators},
   volume={2023},
   ISSN={1029-8479},
   url={http://dx.doi.org/10.1007/JHEP05(2023)215},
   DOI={10.1007/jhep05(2023)215},
   number={5},
   journal={Journal of High Energy Physics},
   publisher={Springer Science and Business Media LLC},
   author={Calò, Simon and Marinissen, Coenraad and Rahn, Rudi},
   year={2023},
   month=may 
}

@article{Rahn:2020eco,
   title={..., 83106786, 114382724, 1509048322, 2343463290, 27410087742, ... efficient Hilbert series for effective theories},
   volume={808},
   ISSN={0370-2693},
   url={http://dx.doi.org/10.1016/j.physletb.2020.135632},
   DOI={10.1016/j.physletb.2020.135632},
   journal={Physics Letters B},
   publisher={Elsevier BV},
   author={Marinissen, Coenraad B. and Rahn, Rudi and Waalewijn, Wouter J.},
   year={2020},
   month=sep, pages={135632} 
}

@article{Babu:2010ex,
  author = {Babu, K. S. and Gabriel, S.},
  title = {Semidirect Product Groups, Vacuum Alignment and Tribimaximal Neutrino Mixing},
  eprint = {1006.0203},
  archivePrefix = {arXiv},
  primaryClass = {hep-ph},
  journal = {Phys. Rev. D},
  volume = {82},
  pages = {073014},
  year = {2010},
  doi = {10.1103/PhysRevD.82.073014}
}

@article{Holthausen:2011pd,
  author = {Holthausen, Martin and Schmidt, Michael A.},
  title = {Natural Vacuum Alignment from Group Theory: The Minimal Case},
  eprint = {1111.1730},
  archivePrefix = {arXiv},
  primaryClass = {hep-ph},
  journal = {JHEP},
  volume = {01},
  pages = {126},
  year = {2012},
  doi = {10.1007/JHEP01(2012)126}
}

\end{document}